\documentclass[usegraphicx,usenatbib,onecolumn]{mn2e}
\usepackage{bm}

\begin{document}
\title{Secondary $B$-mode polarization from Faraday rotation in
  clusters and galaxies}

\author[Tashiro, H. et al.]
{Hiroyuki Tashiro, Nabila Aghanim, and Mathieu Langer\\
  Institut d'Astrophysique Spatiale (IAS), B\^atiment 121, F-91405 Orsay (France);\\
Universit\'e Paris-Sud XI and CNRS (UMR 8617)}

\date{\today}

\maketitle

\begin{abstract}

We revisit the polarisation induced by Faraday rotation when Cosmic
Microwave Background photons traverse magnetised plasma.  We compute
the secondary $B$-mode angular power spectrum from Faraday rotation
due to magnetic fields in galaxies and galaxy clusters with masses
ranging from $10^{11}$ to $10^{16.5} M_\odot$. We investigate its
dependence on the electron and the magnetic field profiles.  Namely,
we consider both the $\beta$-profile of electron density as well as
an electron density distribution based on the Navarro-Frenk-White dark
matter profile. We model the magnetic field structure in galaxies and
clusters motivated by recent observations. We further account for
its redshift evolution and we examine the importance of its coherence length. 
We find that the $B$-mode polarisation from
Faraday rotation depends on the normalisation parameter $C_l\propto
\sigma_8^{5-6}$. At 30~GHz for $\sigma_8=0.8$, the $B$-modes from
Faraday rotation range between $0.01 ~{\mu \rm K}^2$ and $4
\times 10^{-3} ~{\mu \rm K}^2$ at $l=10^4$ in the case of a maximally coherent fields.
For smaller coherence lengths, those amplitudes are smaller and they peak
at higher multipoles.

\end{abstract}


\maketitle

\section{Introduction}

Recent cosmic microwave background (CMB) observations allowed
significant progress in cosmology. The CMB temperature anisotropies
have been measured in detail and, as a result, we know that the
universe is spatially flat and that the power spectrum of the density
fluctuations is consistent with scale-invariance
\citep{spergel-wmap}, strongly supporting the inflation scenario.

One of the main targets for the next generation of CMB observations is
the detailed measurement of the CMB polarisation.  The CMB
polarisation is traditionally split into 2 components: $E$-modes and
$B$-modes \citep{kamiokowski-kosowski-bmode, zaldarriaga-seljak}.
Particularly, the detection of the $B$-mode polarisation is important
for further testing the inflation scenario.  Gravitational waves,
predicted by inflation scenario but not detected yet, generate the
primordial $B$-mode polarisation, while the density fluctuations do
not \citep{kamiokowski-kosowsky, Seljak-Zaldarriaga}.  Therefore, the
measurement of $B$-mode polarisation is an essential step toward
constraining the inflationary scenario.

However there are other sources of $B$-mode polarisation.  One of
them is gravitational lensing \citep{zaldarriaga-seljak}.
Gravitational lensing distorts the CMB polarisation fields so that a
secondary $B$-mode polarisation is induced from the primary
$E$-mode polarisation.  These secondary $B$-modes contaminate the
primary $B$-mode polarisation.  However, significant effort has
been made to find ways how to remove the secondary $B$-mode
polarisation due to gravitational lensing from the polarisation maps
\citep[see for instance][]{hu-okamoto, hirata-seljak}, in order to
recover the primordial $B$-mode angular power spectrum with high
accuracy.
 
The presence of cosmic magnetic fields, observed on various scales, is
also an important source for both primary and secondary $B$-mode
polarisation.  The main ways of detection and measurement of magnetic
fields are Zeeman splitting effect, synchrotron emission and Faraday
rotation.  Zeeman splitting provides direct observation of magnetic
fields. However, the signal is so small with respect to Doppler
broadening that it is essentially useless in the cosmological context
where magnetic fields are week and velocities are high.
Extra-galactic magnetic fields are mainly detected by the other two
methods.  The amplitude of the fields measured in galaxies and galaxy
clusters are typically of the order of $1$--$10$ $\mu$Gauss.  These
methods only provide information on magnetic fields integrated along
the line of sight. Nevertheless, in conjunction with complementary
measurements (electron number density), they have been successfully
applied to show that the coherence length of magnetic fields can be as
large as cluster scales, or even bigger \citep{kim-kromberg}.
Similarly, magnetic fields have been measured with the same amplitudes
in high redshift objects at $z>2$ \citep{athreya-kapahi}.

Although magnetic fields are observed on all scales with increasing
accuracy, we still do not know where they originated from. Primordial
magnetic fields are one of the candidates for the origin of magnetic
fields in galaxies and galaxy clusters.  Primordial magnetic fields
act as sources of the primary $B$-mode polarisation by generating
vorticity in the cosmic plasma and additional gravitational waves
\citep{seshadri-subramanian,subramanian-seshadri,mack-kahniashvili,lewis,tashiro-sugiyama}.
However, those waves arise on small scales and therefore do not
interfere with the detection of the polarisation due to
inflation-generated gravitational waves.

Magnetic fields also create $B$-mode polarisation by Faraday
rotation, i.e. the rotation of the linear polarisation plane through
the interaction between CMB photons and magnetised plasma along their
path.  Faraday rotation distorts the CMB polarisation fields like 
gravitational lensing, so that it induces $B$-mode polarisation from
the $E$-modes.  The effects of Faraday rotation due to primordial
magnetic fields were investigated in some papers
\citep{kosowsky-kahniashvili,scoccola-harari}.
% {check ! \ Miville-Dechenes}. 
In addition to primordial magnetic fields, extra-galactic magnetic
fields of several $\mu$Gauss \citep{carilli-taylor} will produce
Faraday rotation and therefore add to the $B$-mode polarisation.
This effect was first studied by \citet{takada-ohno} who investigated
the $B$-mode polarisation due to Faraday rotation in galaxy
clusters.  They found that homogeneous magnetic fields of several
$\mu$Gauss produce 1 $\mu$K $B$-mode polarisation around $l=1000$.
However, it has been shown that magnetic fields in galaxy clusters are
not uniform (see \citealt{murgia-govoni} and references therein).  The
radial profile of the magnetic fields may affect the $B$-mode
polarisation maps on small scales.  \citet{ohno-takada} discuss the
possibility of the reconstruction of the magnetic fields in a galaxy
cluster from the $B$-mode polarisation map via the
Sunyaev-Zel'dovich (S-Z) effect and X-ray emission.

In this paper, we revisit the secondary $B$-mode polarisation from
Faraday rotation in galaxy clusters. We complete the study by taking
into account galaxies as well. In spiral galaxies, observations
show that the magnetic fields spread widely not only in the galactic
plane but also over the halo, with an average amplitude of about 10
$\mu$Gauss \citep{hummel-beck}.  Even in elliptical galaxies, magnetic
fields corresponding to many $\mu$Gauss are observed
\citep{greenfield-roberts}.  

The paper is organized as follows.  In
Sec.~II, we introduce the formalism of the power spectrum for the
Faraday rotation angle and we derive the angular power spectrum of
$B$-mode polarisation due to Faraday rotation.  In Sec.~III, we
discuss the two different profiles of gas density and magnetic fields
that we use, motivated by observations and numerical simulations.  In
Sec.~IV, the resulting $B$-mode polarisation angular power spectrum
is shown, taking into account the effects of both galaxies and galaxy
clusters.  Sec.~V is devoted to discussions and summary.  Throughout
the paper, we use the WMAP values of cosmological parameters, i.e.,
$h=0.73 \ (H_0=h \times 100 {\rm km/s / Mpc})$, $T_0 = 2.725$K, $h^2
\Omega _{\rm b} =0.0223$ and $h^2 \Omega_{\rm m} =0.128$
\citep{spergel-wmap}.

\section{Polarisation angular power spectrum from Faraday rotation}
\subsection{Angular power spectrum of the rotation angle}

Our aim is to calculate the angular power spectrum of the CMB
polarisation produced by Faraday rotation on the galaxy and
galaxy cluster scales. 
To do so, we first need to compute the angular
power spectrum of the rotation angle.

The rotation angle caused by Faraday rotation is given by
\begin{equation}
\alpha = {2 \pi e^3 \over m_e^2 \omega^2} \int d s ~n_e B  ~\hat {\bm \gamma}
\cdot \hat {\bf b},
\label{eq:faraday-formula}
\end{equation}
where $\omega$ is the angular frequency, $n_e$ is the free electron
number density, $B$ is the magnetic field strength, and $\hat {\bm
\gamma}$ and $\hat {\bf b}$ are the directions of the line of sight
and the magnetic field, respectively.

We then use the halo-formalism commonly used for S-Z power
spectrum computations (e.g. \citet{cole-kaiser, makino-suto,
komatsu-kitayama}). In this formalism, the angular power spectrum is
defined by
\begin{equation}
C_l^\alpha= C_l^{single} +C_l^{halo-halo},
\label{eq:power-rotationmeasure}
\end{equation}
where $C_l^{single}$ describes the Poisson term and $C_l^{halo-halo}$
gives the contribution from correlated halos. They are expressed as
\begin{equation}
C_l ^{\rm single} = \int_0^{z_{\rm dec}} dz \frac{dV}{dz}
\int_{M_{\rm min}}^{M_{\rm max}} dM
{dn(M,z) \over dM} \left|\alpha_l(M,z)\right|^2,
\label{eq:single-power-rot}
\end{equation}
\begin{equation}
C_l ^{\rm halo-halo}=\int_0^{z_{\rm dec}} dz \frac{dV}{dz}
P_{\rm m}\left(k=\frac l{r(z)},z\right)
\left[\int_{M_{\rm min}}^{M_{\rm max}} dM
{dn(M,z) \over dM}b(M,z) \alpha _l(M,z) \right]^2,
\label{eq:halo-power-rot}
\end{equation}
where $r(z)$ and $V(z)$ are the comoving distance and the comoving volume,
respectively.  In Eqs.~(\ref{eq:single-power-rot}) and
(\ref{eq:halo-power-rot}), $n(M,z)$ is the comoving halo number
density of mass $M$ at redshift $z$ and $b(M,z)$ is the linear bias.
We adopt the fitting formula given by \citet{seth-tormen},
\begin{equation}
n(M,z) dM = {\bar \rho \over M} \left(1+2^{-p} {\Gamma(1/2-p)\over  \sqrt \pi} \right)
(1+(q \nu)^{-p})\left({q \nu \over 2 \pi}\right)^{1/2}
\exp \left(- {q \nu \over 2}\right){d \nu \over \nu},
\label{eq:fitting-formula}
\end{equation}
where $\nu = \left( \delta_c \over \sigma(M,z) \right)^2$, $\delta_c$ is the
critical over density and $\sigma$ is the variance smoothed with a
top-hat filter of a scale $R= ( 3M/4 \pi \bar \rho)^{1/3}$ and   
we take $p \approx 0.3$, $q=0.75$ \citep{cooray-sheth}. 
In Eq.~(\ref{eq:halo-power-rot}), we use the linear bias described as
$b(M,z)= 1+ {q \nu -1 /\delta_c} +{2p / \delta_c (1+ (q \nu)^p)}$
\citep{scocciarro-sheth}.

In Eqs.~(\ref{eq:single-power-rot}) and (\ref{eq:halo-power-rot}),
$\alpha_l(M,z)$ is the projected Fourier transform of the rotation
angle obtained in the small angle approximation,
\begin{equation}
\alpha_l = 2 \pi \int d \theta \theta \alpha(\theta,M, z) J_0(l \theta),
\label{eq:angularfourier}
\end{equation}
where $\alpha(\theta,M, z)$ is the angular profile of the rotation
measurement induced by the magnetic field in a galaxy or a galaxy
cluster with mass $M$ at redshift $z$. The rotation angle is obtained
from Eq.~(\ref{eq:faraday-formula}), once a gas distribution and a model
for the magnetic field in clusters are given.
%\ref{eq:beta-model} and \ref{eq:mag-model}. 
We set $\theta=r/D_a$ and $\theta_c = r_c/D_a$ where $D_a$ is the
angular diameter distance.  To compute the power spectrum of the
rotation angle, Eq.~(\ref{eq:power-rotationmeasure}), we also need to
know the angle between the magnetic field and the line of sight.  
In the following, we assume that the orientation of magnetic fields 
does not change within a galaxy or a galaxy cluster,
and we take 
$\left\langle | \hat {\bm \gamma} \cdot \hat{\bf b} | ^2 \right\rangle =1/3$. 
This means that the coherence length of magnetic fields
is the virial radius and 
the orientation of the magnetic fields is random
from structure to structure.
We discuss the case of different coherence lengths
in Sec.~\ref{polarizationpower}.

\subsection{$B$-mode polarisation from Faraday Rotation}

The CMB polarisation can be described using the Stokes parameters $(Q,~U)$.
If CMB photons travel in the $\hat z$ direction,
$Q$ is the difference between the intensity in $\hat y$ and $\hat x$ directions,
and $U$ is the difference between intensities in directions obtained 
by rotating $\hat y$ and $\hat x$ by 45 degrees.
If the CMB radiation passes through a magnetized plasma, it
undergoes Faraday rotation by an angle $\alpha$
(Eq. (\ref{eq:faraday-formula})).
The Stokes parameters after passing through the plasma can be written as
\begin{equation}
\left(
  \begin{array}{c}
    Q'      \\
    U'      \\
  \end{array}
\right)
=\left(
  \begin{array}{cc}
    \cos 2 \alpha  &  \sin 2 \alpha  \\
    - \sin 2 \alpha  &  \cos 2 \alpha  \\
  \end{array}
\right) \left(
  \begin{array}{c}
    Q      \\
    U     \\
  \end{array}
\right).
\end{equation}
Under the assumption $\alpha \ll 1$, we get
\begin{eqnarray}
&&Q' \approx Q + 2 \alpha U,
\nonumber \\
&& U' \approx U - 2 \alpha Q.
\end{eqnarray}
Accordingly, we can write
\begin{equation}
Q' \pm i U' = (Q + 2 \alpha U) \pm i (U - 2 \alpha Q)= (1 \mp 2 i \alpha ) (Q \pm i U).
\label{eq:qu-after}
\end{equation}

The Stokes parameters depend on the choice of the coordinate system.
Therefore it is convenient to introduce the rotation invariant basis,
the $E$-modes and the $B$-modes \citep{kamiokowski-kosowski-bmode,
zaldarriaga-seljak}.  These are obtained by expanding $Q$ and $U$ in
spin-2 spherical harmonics ${}_{\pm 2}Y ^m _l$,
\begin{equation}
Q \pm i U = \sum_{l,m} (E _{lm} \pm i B _{lm}) {}_{\pm 2}Y ^m _l.
\end{equation}
The angle of Faraday rotation can also be decomposed as
\begin{equation}
\alpha = \sum_{l,m} \alpha_{lm} Y^m _l.
\end{equation}
Applying these decompositions to Eq.~(\ref{eq:qu-after}),
we obtain $E' _{lm} \pm i B' _{lm}$ after Faraday rotation as
\begin{eqnarray}
E' _{lm}\pm i B' _{lm} &=& \int d \Omega ~{}_{\pm 2}Y ^{m*} _l
\left(1 \mp 2 i \sum_{l_1,m_1} \alpha_{l_1m_1} Y^{m_1} _{l_1} \right)
\sum_{l_2,m_2} (E _{l_2 m_2} \pm i B _{l_2 m_2}) {}_{\pm 2}Y ^{m_2} _{l_2}
\nonumber \\
&=& E _{l m} \pm i B_{l m}
\mp 2 i \int d \Omega~ {}_{\pm 2}Y ^{m*} _l
 \sum_{l_1,m_1} \alpha_{l_1m_1} Y^{m_1} _{l_1} 
\sum_{l_2,m_2} (E _{l_2 m_2} \pm i B _{l_2 m_2}) {}_{\pm 2}Y ^{m_2} _{l_2}.
\label{eq:fara-eb}
\end{eqnarray}

From this equation, we can get $\Delta B=B'-B$, the contribution of
the Faraday rotation to the power spectrum of $B$-modes.  We give the
detailed calculation in the Appendix and we only write the result
here.  The angular power spectrum of the $B$-mode created by Faraday
rotation is written as
\begin{equation}
C^{\Delta B}_l= N_l^2 \sum_{l_1l_2}
N_{l_2}^2 K(l,l_1,l_2)^2 C^{E}_{l_2}C^\alpha_{l_1}
{(2l_1+1)(2l_2+1)\over 4\pi(2l+1)}\left(C^{l0}_{l_1 0 l_2 0} \right)^2 ,
\label{B--mode-result}
\end{equation}
where $C^{c\gamma}_{a \alpha b \beta}$ are the Clebsch-Gordan coefficients,
$N_l = (2(l-2)!/(l+2)!)^{1/2}$ and 
\begin{equation}
K(l,l_1,l_2)\equiv -{1\over 2}\left(L^2 + L_1^2 + L_2^2 -2L_1L_2
-2L_1L +2L_1-2L_2 -2L\right),
\label{k-factor}
\end{equation}
with $L=l(l+1)$, $L_1=l_1(l_1+1)$, and $L_2=l_2(l_2+1)$.  Preexisting
$B$-mode polarisation acts like a de-polarisation source for
the $B$-modes produced by Faraday rotation (see Appendix,
Eq.~(\ref{eq:b-polar-1})). In Eq.~(\ref{B--mode-result}), we neglect
the contributions of such preexisting $B$-modes created, for example, by
gravitational waves or gravitational lensing. We checked that these
contributions are much smaller (two orders magnitude) than primary
$E$-mode polarisation.

\section{Gas and magnetic field distribution in collapsed objects}

In order to evaluate Faraday rotation on the galaxy or galaxy
cluster scales, we need the distributions of the electron number
density and the magnetic field strength.

\subsection{Gas distribution}

As for the electron number density, we will consider two cases.

\subsubsection{$\beta$-profile}

We first adopt the
$\beta$-profile \citep{cavaliere-fuscofemiano} for both galaxy clusters
and galaxies,
\begin{equation}
n_e(r)=n_{\rm c} \left (1+{r^2  \over r_{\rm c}^2} \right)^{-3 \beta /2},
\label{eq:beta-model}
\end{equation}
where $r$, $r_{\rm c}$ and $n_{\rm c}$ are respectively the physical distance from
the cluster center, the cluster core radius and the central electron
density. We explore a different distribution of the electron number
density in Sec. \ref{sec:NFW}.  Under the
assumption of self-similarity, $r_{\rm c} \propto r_{\rm vir}$,
$r_{\rm vir}$ being the virial radius. Cluster observations
\citep{mohr-mathiesen} roughly give $r_{\rm vir } \approx 10 r_{\rm c}$.
Using the spherical collapse model, the virial radius is related to the
mass $M$ of the dark matter halo and the redshift $z$ through the following
expression, 
\begin{equation}
r_{\rm vir } =\left[ {M \over (4 \pi /3) \Delta_{\rm c}(z) \bar \rho (z) }\right]^{1/3},
\label{eq:virial-radius}
\end{equation}
where $\Delta_{\rm c}(z)=18 \pi^2 \Omega_{\rm m} z^{0.427}$ is the
spherical over density of the virialized halo \citep{nakamura-suto}.

In the $\beta$-profile, the number density of baryons
can be written as
\begin{equation}
\rho_{\rm b}(r)= \mu m_p  n_{\rm c} \left (1+{r^2  \over r_{\rm c}^2} \right)^{-3 \beta /2},
\label{eq:baryon-den}
\end{equation}
where $\mu$ is the mean molecular weight and $m_p$ is the hydrogen
mass.  We can safely assume that most of the halo mass is contained within the virial radius. 
That is, 
\begin{equation}
{\Omega_{\rm b} \over \Omega_{\rm m}} M = 
\int ^{r_{\rm vir}} _0 dr ~ 4 \pi r^2  \rho_{\rm b}(r).
\label{eq:baryon-totalmass}
\end{equation}
From this equation, we can calculate the central electron density
\begin{eqnarray}
n_{\rm c} &= &{\Omega_{\rm b} \over \Omega_{\rm m}} {M \over \mu m_p }  
\left [ {4 \pi \over 3 } r^3 _{\rm vir}~  
{}_2 F_1 (3/2, 3 \beta /2; 5/2; -(r_{\rm vir}/r_{\rm c})^2) \right]^{-1}
\nonumber \\
&=& 9.26 \times 10^{-4} 
\left( {M \over 10^{14} ~M_\odot } \right)
\left( {r_{\rm vir} \over 1~ {\rm Mpc}  } \right)^{-3}
\left( {\Omega_{\rm b} \over \Omega_{\rm m}} \right)
{}_2 F_1 ^{-1} (3/2, 3 \beta /2; 5/2; -(r_{\rm vir}/r_{\rm c})^2)
 ~{\rm cm}^{-3}  ,
\label{eq:baryon-den}
\end{eqnarray}
where $F_1 (\alpha, \beta; \gamma; z)$ is the hypergeometric function.
Using $r_c=10r_{\rm vir}$, we find ${}_2 F_1 (3/2, 3 \beta /2; 5/2;
-(r_{\rm vir}/r_{\rm c})^2)$ is 0.032 for $\beta = 0.6$ and 0.005 for
$\beta = 1.0$.

\subsubsection{Navarro-Frenk-White profile}\label{sec:NFW}

We also consider a different electron density profile from the
$\beta$-profile. Namely, we use a density profile motivated by the
Navarro-Frenk-White (NFW) dark matter profile \citep{navarro-frenk}.
  
The NFW dark matter density profile is given by:
\begin{equation}
\rho_{\rm dm}(x) = \frac{\rho_{\rm s}}{x(1+x)^2}.
\label{eq:NFWprofiles}
\end{equation}
Here $x \equiv r/r_{\rm s}$ where $r_{\rm s}$ is a scale radius, and
$\rho_{\rm s}$ is a scale density at this radius.  The scale radius,
$r_{\rm s}$, is related to the virial radius
\begin{equation}
{r_{\rm s}(M,z)}={r_{\rm vir}(M,z) \over c(M,z)},  
\label{eq:scaleradius}
\end{equation}
where $c$ is the concentration parameter. In the following, we adopt
the concentration parameter of \citet{komatsu-seljak},
\begin{equation}
c \approx \frac{10}{1+z}\left[\frac{M}{M_*(0)}\right]^{-0.2},
\label{eq:concentrait}
\end{equation}
where $M_*(0)$ is a solution to $\sigma(M)=\delta_c$ at the redshift
$z=0$.

The electron number density profile $n_e$ can then be obtained from
the NFW dark matter profile following \citet{komatsu-seljak}. For
this, three assumptions are made: the gas is in hydrostatic
equilibrium in the dark matter potential, the gas density follows the
dark matter density in the outer parts of the halo, and finally the
equation of state of the gas is polytropic, $P_{\rm gas}\propto \rho_{\rm
gas}^{\gamma}$ (where $P_{\rm gas}$, $\rho_{\rm gas}$ and ${\gamma}$
are the gas pressure, the gas density and the polytropic index).  From
these assumptions, with $n_e \propto \rho_{\rm gas} /\mu
m_p$, we obtain the electron number density profile
\begin{equation}
n_e(x)= n_c
\left\{
1 - F\left[1-\frac{\ln(1+x)}x\right]
\right\}^{1/\left(\gamma-1\right)},
\label{eq:gasprofile}
\end{equation}
where the central electron density $n_{\rm c}$ and the coefficient $F$ are
\begin{equation}
n_{\rm c}
=3.01 
\left( {M \over 10^{14} M_\odot } \right)
\left( {r_{\rm vir} \over 1~ {\rm Mpc}  } \right)^{-3}
\left( {\Omega_{\rm b} \over \Omega_{\rm m}} \right)
\frac{c}{(1+c)^2}
\left[\ln(1+c)-\frac{c}{1+c}\right]^{-1}
\left[{c \over c - F [c-\ln(1+c)] }\right]^{1/(\gamma-1)}
~{\rm cm}^{-3}     ,
\label{eq:rhog0}
\end{equation}
\begin{equation}
F\equiv3\eta^{-1}_{\rm c}\frac{\gamma-1}{\gamma}
\left[\frac{\ln(1+c)}c-\frac1{1+c}\right]^{-1}.
\label{eq:Bcoefficient}  
\end{equation}
\citet{komatsu-seljak} provide the following useful fitting formulae for
$\gamma$ and $\eta_{\rm c}$:
\begin{equation}
\gamma= 1.137 + 8.94\times 10^{-2}\ln(c/5) - 3.68\times 10^{-3}(c-5),
\label{eq:gammafitting}
\end{equation}
\begin{equation}
\eta_{\rm c}= 2.235 + 0.202(c-5) - 1.16\times 10^{-3}\left(c-5\right)^2.
\label{eq:etafitting}
\end{equation}

\subsection{Magnetic field distribution}

Magnetic fields in galaxies can reach amplitudes up to 10 $\mu$Gauss
\citep{hummel-beck}.  Observations by \citet{murgia-govoni}
suggest that the distribution of
magnetic fields in galaxies and galaxy clusters is such that their strength
decreases outward.  
Therefore, in the $\beta$-profile,
we adopt the form proposed by those authors:
\begin{equation}
B(r)= B_{\rm c} \left(1+ {r^2 \over r_{\rm c}^2} \right)^{-3\beta \mu/2 },
\label{eq:mag-model}
\end{equation}
where $B_{\rm c}$ is the mean magnetic field strength at the center,
$\beta$ is the parameter of the $\beta$-profile.  From
Eq.~(\ref{eq:beta-model}), we get $B \propto n_e ^{\mu\beta}$. When
$\mu=1 /2$, the energy density of magnetic fields decreases outward
like the electron number density, while in the case where $\mu=2 /3$,
magnetic fields are frozen into matter.

In the NFW case, the distribution of magnetic fields
in galaxies and clusters is given by:
\begin{equation}
B(x)= B_{\rm c}
\left\{
1 - F\left[1-\frac{\ln(1+x)}x\right]\right\}^{\mu/\left(\gamma-1\right)}.
\label{eq:bpara-profile}
\end{equation}
If $\mu=2 /3$, magnetic fields are frozen into the matter whereas, if
$\mu=1/2$, the energy density of magnetic fields decreases from the
center like the electron number density.  

We further model the growth of magnetic fields in galaxies and galaxy
clusters, considering the dynamo process which is widely, although not
unanimously, accepted \citep{widrow}. In this scenario, magnetic
fields are amplified from weak seeds, most likely produced by
astrophysical processes, by the dynamo process and the time scale of
the amplification is of the order of the dynamical time scale, $t_{\rm
d} = \sqrt{ r_{\rm vir } ^3 / G M}$.  Assuming typical magnetic field
amplitudes of 3 and 10 $\mu$Gauss for clusters and galaxies
respectively, we model the magnetic field growth by
\begin{equation}
B_{\rm c}= \left\{
\begin{array}{cc}
    3\exp \left(- (t_0-t(z)) /t_{\rm d}\right) ~{ \mu \rm Gauss}, 
    & 10^{13} ~M_\odot \le M < 10^{16.5} ~M_\odot,  \\
     10 \exp \left(- (t_0-t(z)) /t_{\rm d}\right) ~{ \mu \rm Gauss},
    & 10^{11} ~M_\odot \le M < 10^{13} ~M_\odot,  \\
\end{array}
\right.
\label{eq:magneticamplitude}
\end{equation}
where $t_0$ is the present time.

\section{Results}

In order to evaluate the effect of Faraday rotation due to
galaxies and galaxy clusters on CMB polarisation, we arbitrarily
consider that the contribution from galaxies is associated with
halos of mass lower than $10^{13} ~M_\odot$. Objects with halo masses
larger than this limit are considered as galaxy clusters. The
particle number density in galaxies is higher than that in galaxy
clusters. Moreover, in cold dark matter scenario, the number density
of galaxies is higher than that of galaxy clusters.  Therefore the
Faraday rotation produced in galaxies presumably cannot be neglected
against that generated in galaxy clusters.

\subsection{Rotation angle power spectrum}

\begin{figure}
  \begin{center}
    \includegraphics[keepaspectratio=true,height=50mm]{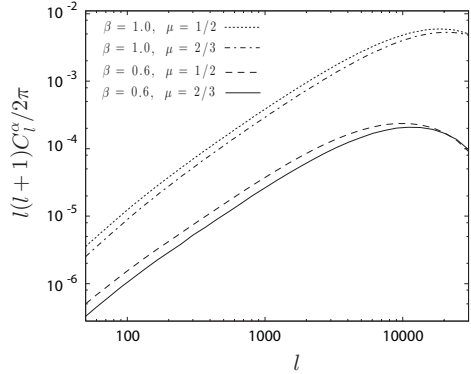}
  \end{center}
  \caption{The power spectra of Faraday
  rotation angle caused by $3~\mu$Gauss magnetic fields in galaxy
  clusters for the $\beta$-profile case. The
  solid line is for $\beta =0.6$ and $\mu=2/3$ and the dashed line is
  for $\beta =0.6$ and $\mu=1/2$. We also plot the case of $\beta
  =1.0$ and $\mu=1/2$, and $\beta =1.0$ and $\mu=2/3$ as a dotted line
  and a dashed-dotted line, respectively.  We use $\sigma_8 = 0.8$
  and set the CMB frequency to 30~GHz.}
  \label{fig:rot-beta}
\end{figure}

\begin{figure}
  \begin{center}
    \includegraphics[keepaspectratio=true,height=50mm]{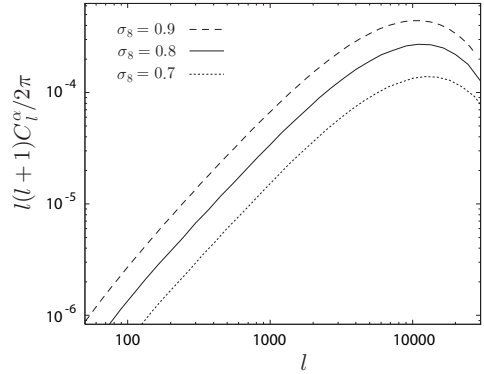}
  \end{center}
  \caption{The dependence on the density fluctuation amplitude of the
  power spectra of Faraday rotation angle. We choose 30 GHz as the CMB
  frequency and adopt the $\beta$-profile with $\beta =0.6$,
  $\mu=2/3$ and  $3~\mu$Gauss magnetic fields. The dashed, the solid,
  the dotted lines are $\sigma=0.9$, $\sigma=0.8$ and $\sigma=0.7$,
  respectively.}
  \label{fig:rot-betasigma}
\end{figure}

In Fig.~\ref{fig:rot-beta}, we show the angular power spectrum of the
Faraday rotation angle due to galaxy clusters only.  We assume that
their mass range is between $10^{13} ~M_{\odot}$ and $10^{16.5}
~M_{\odot}$ and that all clusters follow the $\beta$-profile with a
magnetic field distributed following Eq.~(\ref{eq:mag-model}) and
$B_{\rm c}=3~\mu {\rm G}$. We checked that the Poisson term in
Eq.~(\ref{eq:power-rotationmeasure}) dominates in the mass region of galaxy
clusters. The contribution from correlated halos is thus neglected.
We note that increasing $\beta$ increases the amplitude of the power
spectrum and shifts its peak to smaller angular scales. This is
because large $\beta$ value gives steeper profiles and smaller core
radii.  The magnetic field profile also affects the angular power
spectrum. However, this dependence is weaker than that of the electron
density profile.  Fig.~\ref{fig:rot-betasigma} shows the dependence of
the angular power spectrum on the normalisation of the density
fluctuation $\sigma_8$.  We find $l^2 C_l \propto \sigma_8 ^5$
somewhat different from the Sunyaev-Zel'dovich (S-Z) case
\citep{komatsu-seljak}. We compared this result with that of
\citet{takada-ohno} and found it in general agreement. The location of the
peaks in our case and their case is slightly different. This can be
explained by the fact that we take into account distribution of the
magnetic fields with central peak, while \citet{takada-ohno} consider
homogeneous magnetic fields over the whole cluster scale.

\begin{figure}
  \begin{center}
    \includegraphics[keepaspectratio=true,height=50mm]{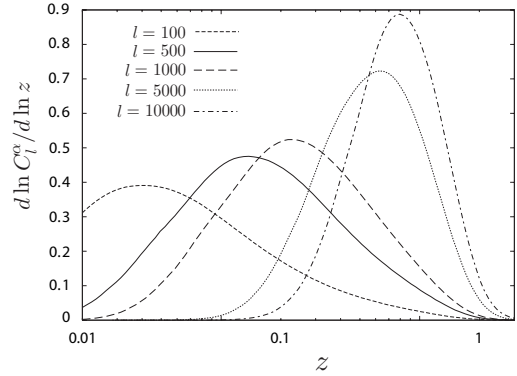}
  \end{center}
  \caption{The redshift contribution for various $l$ modes.  The
  short-dashed, the solid, the long-dashed and the dashed dotted
  lines represent $l=100$, $l=500$, $l=1000$, $l=5000$ and $l=10000$,
  respectively.  This figure is calculated for the $\beta$-profile with
  $\beta =0.6$ and $\mu=2/3$ and $3~\mu$Gauss magnetic fields.}
  \label{fig:red-beta}
\end{figure}

\begin{figure}
  \begin{center}
    \includegraphics[keepaspectratio=true,height=50mm]{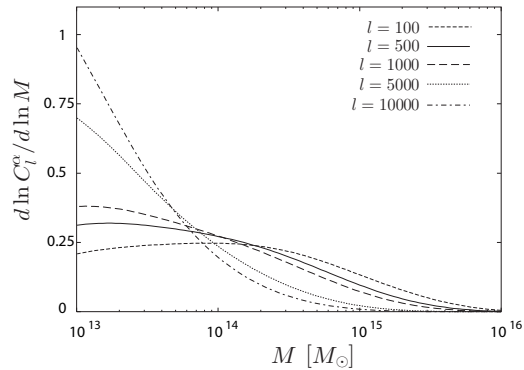}
  \end{center}
  \caption{The mass contribution for various $l$ modes.  The
  short-dashed, the solid, the long-dashed and the dashed dotted
  lines represent $l=100$, $l=500$, $l=1000$, $l=5000$ and $l=10000$,
  respectively.  This figure is calculated for the $\beta$-profile with
  $\beta =0.6$ and $\mu=2/3$ and $3~\mu$Gauss magnetic fields.}
  \label{fig:mass-beta}
\end{figure}

In Fig.~\ref{fig:red-beta}, we represent the redshift distribution of
$C_l$, ${d \ln C_l / d \ln z}$, for different $l$s.  The power
spectrum of the rotation angle on large scales (small $l$) is mostly
due to lower redshift galaxy clusters.  The rotation angle strongly
depends on the frequency (see Eq.~(\ref{eq:faraday-formula})). As a
result, low redshift clusters have a larger contribution to $C_l$
than in the S-Z effect.  For example, while most of the contribution
at $l=10000$ in the S-Z power spectrum comes from galaxy clusters at
redshift $z=2$ \citep{komatsu-seljak}, the power spectrum of the
Faraday rotation angle at the same multipole is associated with galaxy
clusters at redshift $z=1$. On scales below $l=100$, the contribution
to the rotation angle $C_l$ from galaxy clusters at redshift lower
than $z=0.01$ is large. The angular distance of the redshift $z=0.01$
is about $40$ Mpc while the virial radius of the galaxy cluster with
mass $10^{13} ~M_{\odot}$ is about $2$ Mpc.  This means that the small
angle approximation used in Eq.~(\ref{eq:angularfourier}) is
not valid at $l<100$. Similarly, we plot the mass distribution of
$C_l$ for different $l$s, ${d \ln C_l / d \ln M}$, in
Fig.~\ref{fig:mass-beta}.  For all $l$s, the main contribution is due
to galaxy clusters with masses between $10^{13}~M_{\odot}$ and
$10^{14} ~M_{\odot}$.  The amplitude of the rotation angle power
spectrum at multipoles larger than $l=5000$ is mostly due to clusters
with $10^{13}~M_\odot$.  Comparing the mass contribution of the S-Z
effect \citep{komatsu-seljak}, we find that the Faraday rotation angle
is less affected by the massive halos.  This is because the
high-redshift clusters do not contribute significantly to the
amplitude of the Faraday rotation spectrum (see discussion on the
redshift dependence above).

\par\bigskip

\begin{figure}
  \begin{center}
    \includegraphics[keepaspectratio=true,height=50mm]{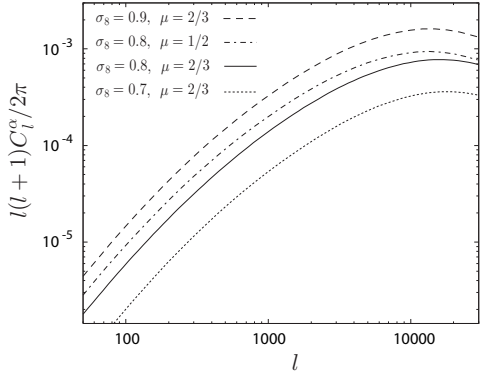}
  \end{center}
  \caption{The angular power spectra of Faraday rotation angle caused
  by galaxy clusters and galaxies.  The solid line is the case of the
  $\beta$-profile with $\beta =0.6$ and $\mu=2/3$ and the
  dashed-dotted line is the case of the $\beta$-profile with $\beta
  =0.6$ and $\mu=1/2$. Both are calculated using $\sigma_8=0.8$.  The
  dashed and the dotted lines are plotted as the case of
  $\sigma_8=0.9$ and $\sigma_8=0.7$, respectively. In all plots, the
  CMB frequency is 30 GHz.}
  \label{fig:rot-glxbeta}
\end{figure}

\begin{figure}
  \begin{center}
    \includegraphics[keepaspectratio=true,height=50mm]{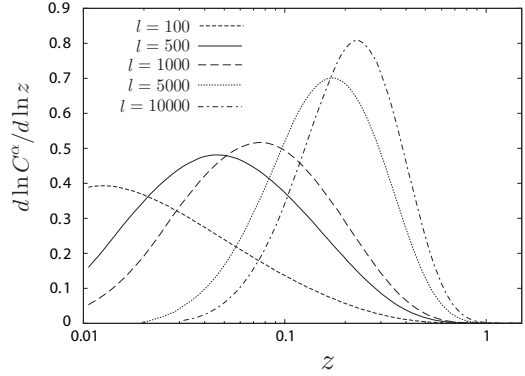}
  \end{center}
  \caption{The redshift contribution to various $l$ modes.  The short
  dashed the solid, the long dashed and the dashed dotted lines
  represents $l=100$, $l=500$, $l=1000$, $l=5000$ and $l=10000$,
  respectively.  This figure is calculated for galaxies and galxy clusters with the $\beta$-profile for
  $\beta =0.6$ and $\mu=2/3$.}
  \label{fig:red-glxbeta}
\end{figure}

\begin{figure}
  \begin{tabular}{cc}
   \begin{minipage}{0.5\textwidth}
  \begin{center}
    \includegraphics[keepaspectratio=true,height=50mm]{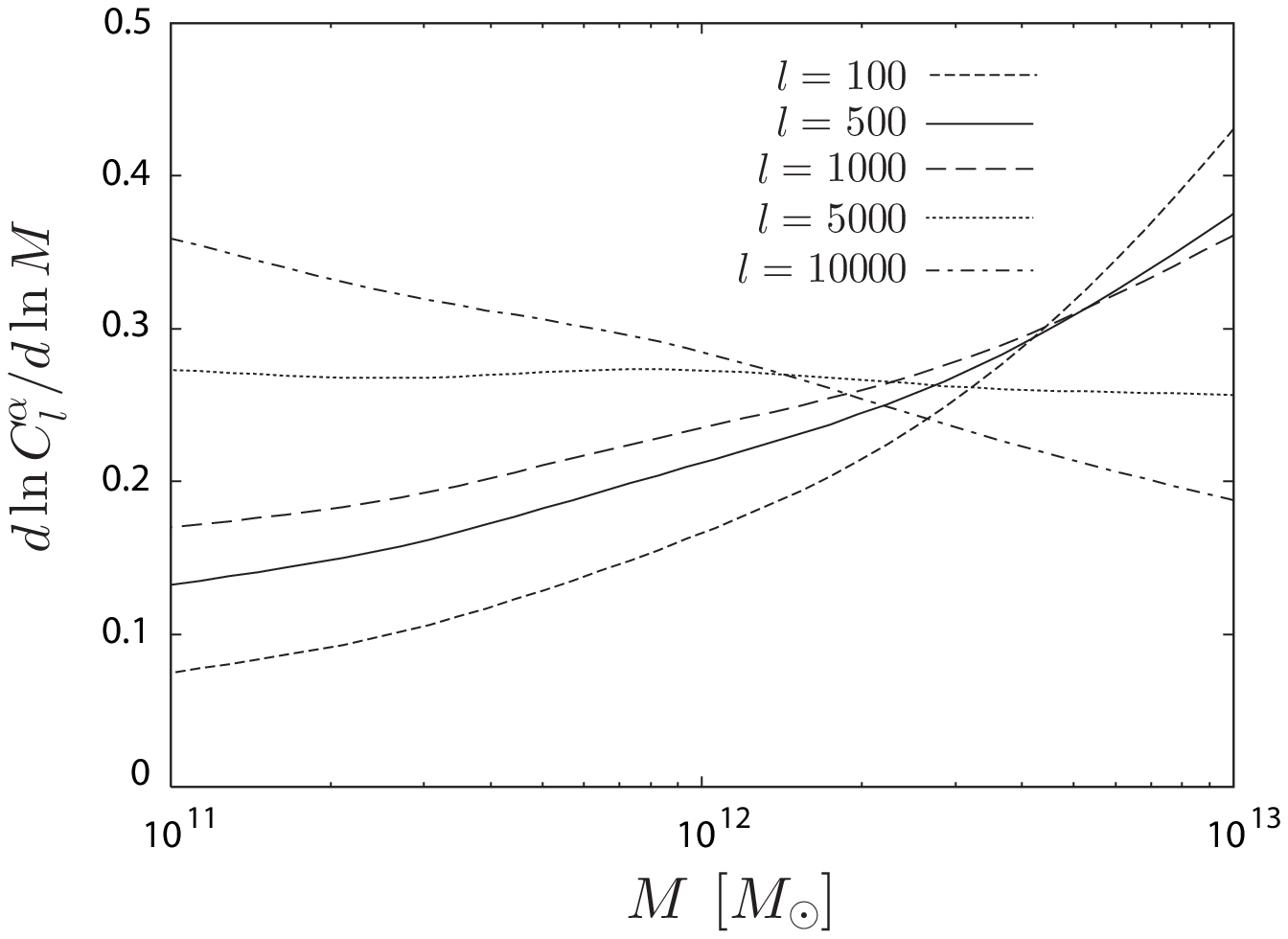}
  \end{center}

  \end{minipage}
   \begin{minipage}{0.5\textwidth}
  \begin{center}
    \includegraphics[keepaspectratio=true,height=50mm]{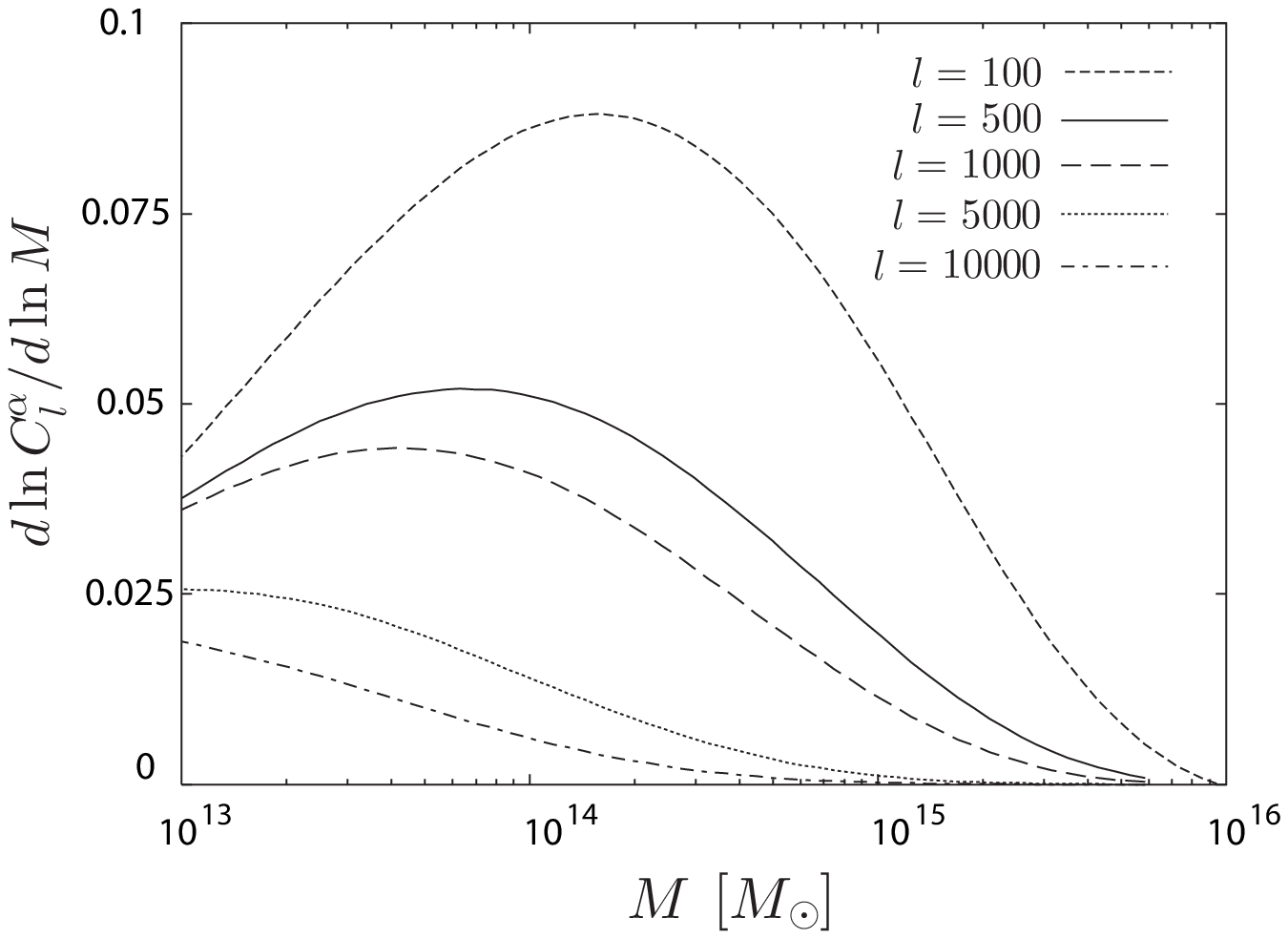}
  \end{center}
%  \caption{The mass contribution to various $l$ modes.
%  The short dashed the solid, the long dashed and the dashed dotted lines 
%  represents $l=100$, $l=500$, $l=1000$, $l=5000$ and $l=10000$, respectively.
%  This figure is calculated in the NFW profile with $\mu=2/3$.}
%  \label{fig:mass-glxbetacluster}
   \end{minipage}
  \end{tabular}
  \caption{The mass contribution to various $l$ modes.
  The left panel shows the contribution from the galaxy mass range
  while the right panel shows that of the galaxy cluster mass range.
  The short dashed the solid, the long dashed and the dashed dotted lines 
  represent $l=100$, $l=500$, $l=1000$, $l=5000$ and $l=10000$, respectively.
  This figure is calculated for galaxies and galaxy cluster with the $\beta$-profile 
  for $\beta =0.6$ and $\mu=2/3$.}
  \label{fig:mass-glxbeta}
 
\end{figure}

If we take into account the contribution from the whole population of
structure including galaxy clusters and galaxies, we obtain the results
shown in Fig.~\ref{fig:rot-glxbeta}. We calculate both Poisson and
halo-halo correlation terms in Eq.~(\ref{eq:power-rotationmeasure}) in
the mass range $10^{11} ~M_\odot \le M < 10^{13} ~M_\odot$. In the mass
range of clusters $10^{13} ~M_\odot \le M < 10^{16.5} ~M_\odot$, we
neglect the halo-halo correlation term as mentioned previously. The
peak of the power spectrum is shifted to smaller angular scales
($l=20000$) because of the contribution from galaxies.  In
Fig.~\ref{fig:rot-glxbeta}, we use $\beta=0.6$ and check that varying
$\mu$ in Eq.~(\ref{eq:mag-model}) does not modify significantly the
power spectrum. We also compare the results with different
normalisation parameters. The dependence of the amplitude of the
rotation measurement is approximately $l^2 C_l \propto \sigma_8 ^{5.5
\sim 6 }$.  The redshift distribution for the power spectrum computed
using the $\beta$-profile is given in Fig.~\ref{fig:red-glxbeta}. It
also takes into account the evolution of magnetic field
Eq.~(\ref{eq:magneticamplitude}) which influences the redshift
distribution. The contribution of high redshift objects (mainly
galaxies) is suppressed for $z>1$. Moreover the redshift distribution
at high multipoles is shifted to lower $z$ (see dot-dashed line in
Fig.~\ref{fig:red-glxbeta}). Most of the Faraday rotation-induced
signal is thus due to low redshift objects. If we now explore the mass
distribution of $C_l$ for different $l$ (Fig.~\ref{fig:mass-glxbeta}),
we notice that the contribution from galaxies (left panel) overwhelms
that of galaxy clusters (right panel). We also note the difference in
amplitudes between the two panels is mainly due to the difference in magnetic
field strength between galaxy clusters (3$~\mu$G) and galaxies
(10$~\mu$G) described in Eq.~(\ref{eq:magneticamplitude}).

\par\bigskip

\begin{figure}
  \begin{center}
    \includegraphics[keepaspectratio=true,height=50mm]{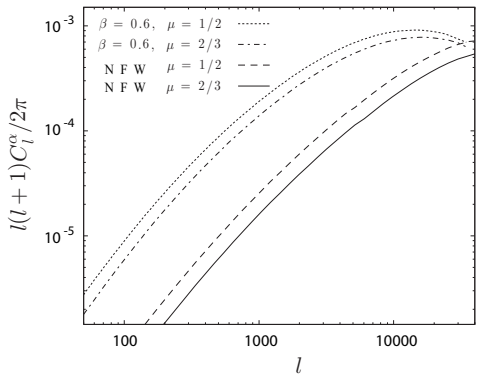}
  \end{center}
  \caption{The angular power spectra of Faraday rotation angle by
  galaxy clusters and galaxies at $30$ GHz CMB frequency. The solid
  line is the NFW case with magnetic fields frozen into matter,
  $\mu=2/3$, and the dashed line is the NFW case with magnetic fields
  which energy density decrease outward like the electron
  distribution.  We also plot the $\beta$-profile cases with
  $\mu=2/3$ and $\mu=1/2$ as the dashed-dotted and the dotted lines,
  respectively.  }
  \label{fig:rot-nfwglx}
\end{figure}

Now, following the same approach as for the $\beta$-profile, we compute
the angular power spectrum of Faraday rotation angle for the NFW
profile given in Fig.~\ref{fig:rot-nfwglx}.  In this plot, we show the
results at 30 GHz, and we assume that the magnetic field distribution
within clusters or galaxies writes as Eq.~(\ref{eq:bpara-profile}).  For
comparison, we plot in the same figure the angular power spectrum for the
$\beta$-profile and magnetic field given by Eq.~(\ref{eq:mag-model}).  The
NFW profile is more peaked than the $\beta$-profile. As a result, the
power spectrum using the NFW profile is shifted to smaller scale
(multipoles larger than 30000) as compared to those of $\beta$-profile.

\subsection{Polarisation power spectrum}\label{polarizationpower}

\begin{figure}
  \begin{center}
    \includegraphics[keepaspectratio=true,height=50mm]{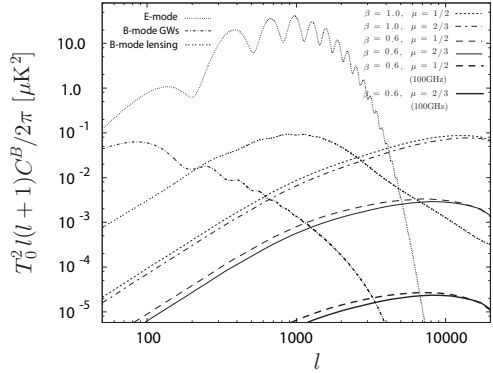}
  \end{center}
  \caption{The angular power spectra of the $B$-mode CMB polarisation
  caused by Faraday rotation for the $\beta$-profile.  The solid line
  is for $\beta=0.6$ and $\mu=2/3$, the dashed line is for $\beta=0.6$
  and $\mu=1/2$, the dashed-dotted line is $\beta=0.6$ and $\mu=2/3$, and
  the dotted line is for $\beta=1.0$ and $\mu=1/2$.  We use
  $\sigma_8=0.8$ and set the CMB frequency to $30~$GHz. For
  comparison, we plot, for 100~GHz, the spectrum for $\beta=0.6$ and
  $\mu=2/3$ (thick solid line), and for $\beta=0.6$ and $\mu=2/3$
  (thick dashed line).  We also give the power spectra of the
  primordial $E$-mode polarisation, the $B$-mode polarisation caused
  by gravitational waves with a tensor to scalar ratio $r=0.6$, and the
  gravitational lensing $B$-mode polarisation.  }
  \label{fig:cl-beta}
\end{figure}

We show, in Fig.~\ref{fig:cl-beta}, the angular power spectrum of the
$B$-mode polarisation caused by Faraday rotation at 30 GHz using
a $\beta$-profile for the gas distribution.  From
Eq.~(\ref{B--mode-result}), the $B$-mode polarisation by Faraday
rotation depends on both the primordial $E$-mode polarisation $C_l
^E$ and the rotation power spectrum $C_l ^\alpha$.  However, since the
rotation power spectrum peaks on smaller angular scales than the
primordial $E$-mode polarisation, the produced $B$-mode polarisation
reflects the rotation angle characteristics.  They both peak around
$l\sim8000$ and for a $\beta$-profile the peak value as a function the
main parameters of the model is
\begin{equation}
{l(l+1)  \over 2 \pi} C_l ^B \approx 3.0 \times 10^{-3}  
\left( {B_{\rm c} \over 3~\mu {\rm Gauss}} \right)^2
\left( {\sigma_8 \over 0.8} \right)^5
\left( {\nu \over 30 ~{\rm GHz} } \right)^{-4} ~{\mu \rm K}^2.
\label{eq:approx-clbeta}
\end{equation}
We also give the power spectrum of $B$-mode from Faraday rotation at
100 GHz (thick lines) which is the optimal band for CMB observation.

For comparison, we plot in the same figure the power spectrum of the
primary $E$-mode polarisation (thin dotted line), the $B$-mode
polarisation due to the gravitational waves (thin dot-dashed line) and the
lensing $B$-mode polarisation (thin dot-doted line).  The polarisation
from gravitational waves is computed with a tensor to scalar ratio
$r=0.6$ which is the upper limit from the WMAP three years data
\citep{spergel-wmap}.  This signal dominates on large angular
scales. On small scales, the power spectrum of the polarisation caused
by Faraday rotation dominates the $B$-mode polarisation due to
gravitational lensing at frequencies lower than 30 GHz, for reasonable
values of the cluster magnetic field. When the frequency increases,
the gravitational lensing $B$-mode polarisation dominates.  We plot
the power spectrum of the Faraday induced polarisation for different
parameters $\beta$ and $\mu$ describing the electron density and the
magnetic field profiles (see line styles in caption). Similarly to the
rotation angle power spectrum Fig. \ref{fig:rot-beta}, we note that
varying $\mu$ has very little effect whereas varying $\beta$ changes
the amplitude of the power spectrum by two orders of magnitude.
 
\begin{figure}
  \begin{center}
    \includegraphics[keepaspectratio=true,height=50mm]{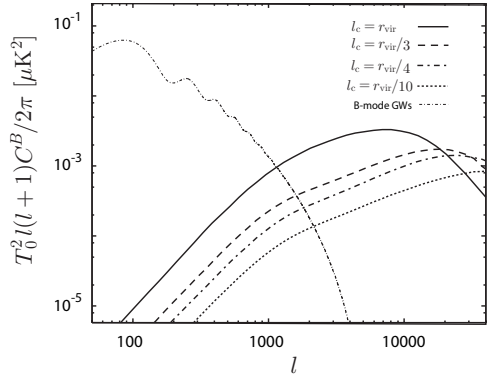}
  \end{center}
  \caption{
  The angular power spectra of the $B$-mode CMB polarisation
  by Faraday rotation for the different coherence lengths.  The solid line
  corresponds to the model with $l_{\rm c} = r_{\rm vir}$.
  The dashed, the dot-dashed and the dotted lines are 
  for $l_{\rm c} = r_{\rm vir}/3$, $l_{\rm c} = r_{\rm vir}/4$ 
  and $l_{\rm c} = r_{\rm vir}/10$.
  For comparison, we plot the $B$-mode polarisation caused
  by gravitational waves.
  }
  \label{fig:coherent}
\end{figure}

In the above calculation,
we assumed that the coherence length of the magnetic field in a galaxy cluster
is the virial radius.
This means that the inner product of the directions of the line of sight
and of the magnetic field in Eq.~(\ref{eq:faraday-formula}) is constant over a galaxy cluster.
However, many observations suggest that the coherence length is shorter than the virial radius
(e.g. \citet{Brandenburg-Subramanian}). 
In order to evaluate the effect of different coherence lengths, 
we assume that the direction of the magnetic field in a galaxy cluster 
changes smoothly and the inner product of directions in Eq.~(\ref{eq:faraday-formula})
has the following form,
\begin{equation}
\hat {\bm \gamma} \cdot \hat {\bf b} = \cos \left( {s \over l_{\rm c}} \pi \right),
\end{equation}
where $l_{\rm c}$ is the coherence length.
Taking $l_{\rm c}= r_{\rm vir}/3$, $r_{\rm vir}/4$ and $r_{\rm vir}/10$,
we calculate the angular power spectrum of $B$-mode polarisation
and plot the results in Fig.~\ref{fig:coherent}.  
The small coherence length implies many changes of the direction of magnetic fields 
so that depolarisation occurs along the line of sight.
When the coherence length is one tenth of the virial radius,
the amplitude of the angular spectrum is decreased
and the peak position shifts to small scales.

\begin{figure}
  \begin{center}
    \includegraphics[keepaspectratio=true,height=50mm]{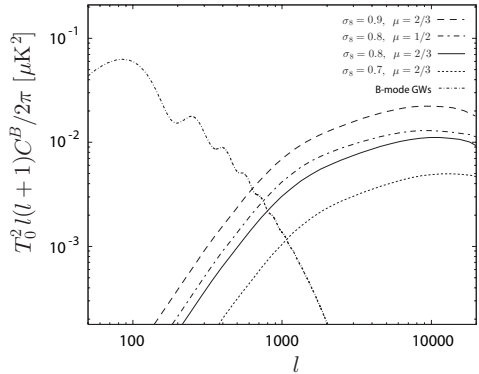}
  \end{center}
  \caption{The angular power spectra of the $B$-mode polarisation
  caused by Faraday rotation in galaxy clusters and galaxies for the
  $\beta$-profile.  The solid line is for $\beta =0.6$ and $\mu=2/3$,
  and the dashed-dotted line is for $\beta =0.6$ and $\mu=1/2$.  We
  also plot the case of $\sigma_8=0.9$ and $\sigma_8=0.7$ for $\beta
  =0.6$ and $\mu=2/3$ (dashed and dotted lines respectively).  For
  comparison, we give the $B$-mode polarisation induced by
  gravitational waves with $r=0.6$.}
  \label{fig:cl-glxbeta}
\end{figure}

Next, the angular power spectrum of $B$-polarisation caused by
Faraday rotation in galaxy clusters and galaxies is plotted in
Fig.~\ref{fig:cl-glxbeta}.  We plot this figure for a
$\beta$-profile with $\beta=0.6$ and $\mu=2/3$, and at a frequency of
30 GHz.  The power spectrum peaks around $l=10000$ with a peak
amplitude, which depends on $\sigma_8$, $\nu$ and $B_c$,
\begin{equation}
{l (l+1)  \over  2 \pi} C_l ^B \approx 1.1 \times 10^{-2}
\left( {B_{\rm c} \over 10~\mu {\rm Gauss}} \right)^2
\left( {\sigma_8 \over 0.8} \right)^6
\left( {\nu \over 30 ~{\rm GHz} } \right)^{-4} ~{\mu \rm K}^2.
\label{eq:approx-clglx}
\end{equation}
The $B$-mode polarisation induced by galaxies can be the major
component of $B$-mode polarisation on small scales. If the average
magnetic field of galaxies is $10~\mu$Gauss, the Faraday rotation
produces the dominant $B$-mode polarisation at $30$ GHz for
$l>4000$. Alternatively, on scales smaller than $l=10000$, the
$B$-mode polarisation generated by galaxies for the same magnetic
field is expected to dominate at frequencies up to 60 GHz.
 
\begin{figure}
  \begin{tabular}{cc}
   \begin{minipage}{0.5\textwidth}
  \begin{center}
    \includegraphics[keepaspectratio=true,height=45mm]{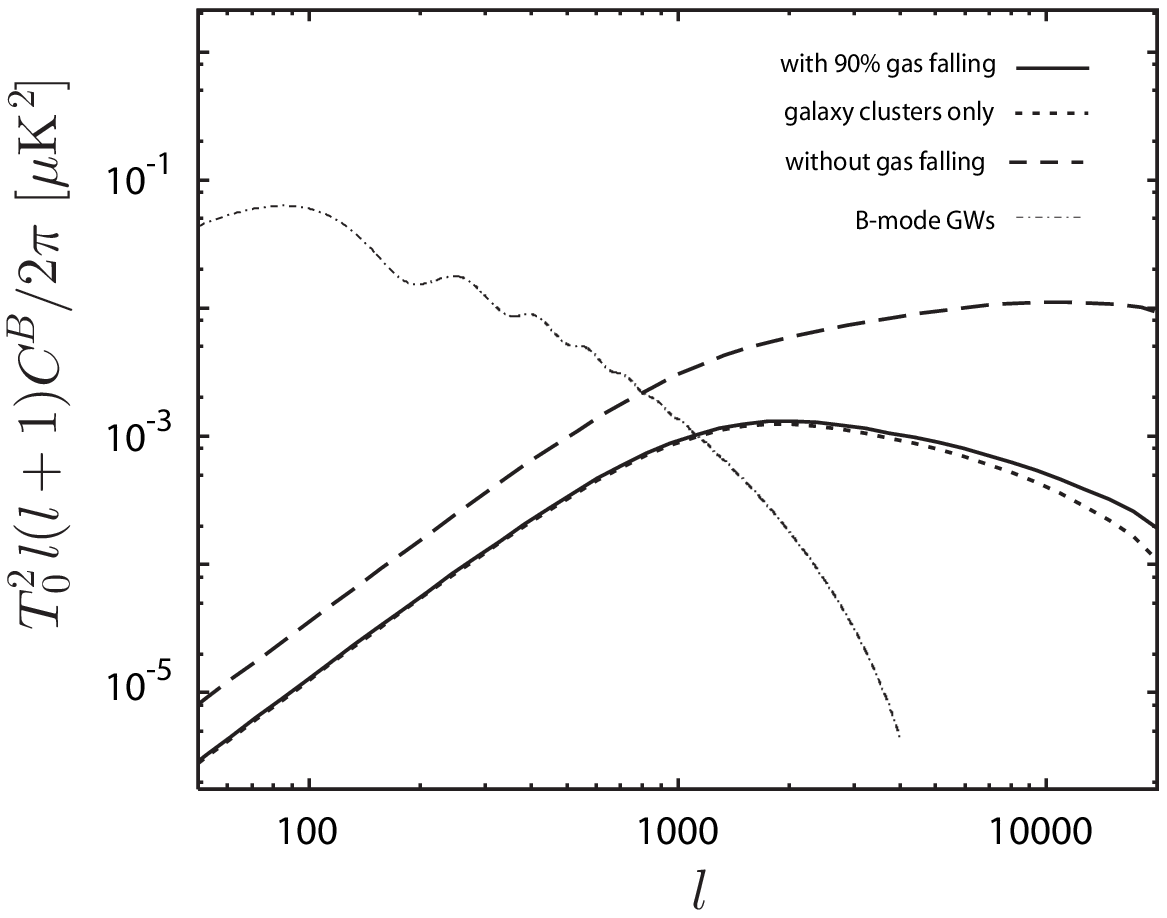}
 \end{center}
  \end{minipage}
     \begin{minipage}{0.5\textwidth}
  \begin{center}
    \includegraphics[keepaspectratio=true,height=45mm]{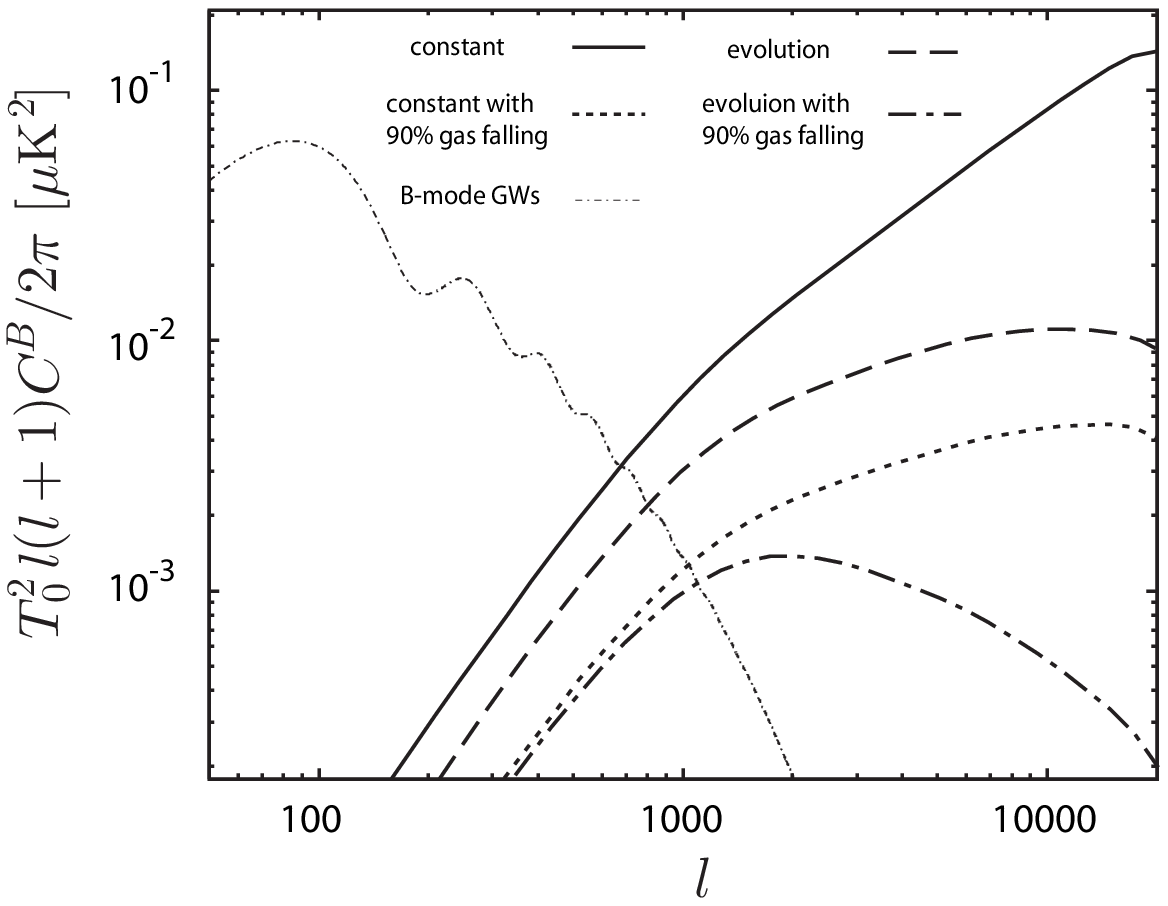}
  \end{center}
   \end{minipage}
  \end{tabular}
    \caption{(the left panel) The angular power spectra of the $B$-mode polarisation
  caused by Faraday rotation in galaxy clusters and galaxies.
  For the solid line, we assume that 90 \% of gas in galaxies are condensed
  in the disk. The dotted line shows the contribution from galaxy clusters only.
  For comparison, we plot the power spectrum for the case without gas depletion and that 
  due to gravitational waves with $r=0.6$ 
  as the dashed and the dashed-dotted lines, respectively.
  (the right panel) The angular power spectra of the $B$-mode polarisation
  for different evolutions of magnetic fields.  
  The solid line is for the model where magnetic fields
  are constant. The dotted line corresponds to the case where magnetic fields are constant
  and gas is depleted in galactic halos.  We
  also plot the case where the time scale of the magnetic field evolution is 
  the dynamical timescale as the dashed line.
  The dashed-dotted line is takes into account both magnetic field evolution 
  and gas depletion.}
  \label{fig:noevoglx}

\end{figure}

As a galaxy evolves, gas is falling from the halo onto the disk
and the halo gas density decreases.
The remaining fraction of diffuse baryons in the galaxy halo at low redshift can be as low as 10 \%,
whereas in a galaxy cluster it roughly stays equal to 1 \citep{binney-galactic}. 
The evolution of the gas in a galaxy is expected to affect the power spectrum.
We estimate the effect of the gas depletion by assuming that
only one tenth of the total baryon density is left in the halo and 
the rest is condensated on the galactic disk, so that
\begin{equation}
\alpha = {2 \pi e^3 \over m_e^2 \omega^2} \int d s ~ f_g n_e B  ~\hat {\bm \gamma}
\cdot \hat {\bf b},
\label{eq:faraday-formula2}
\end{equation}
where
\begin{equation}
f_g = 
\left\{
  \begin{array}{cc}
    0.1,   &  M<10^{14} M_\odot,   \\
    1,  &   M>10^{14} M_\odot. \\
  \end{array}
\right.
\label{eq:laa}
\end{equation}
Ninety percent depletion is the maximum we can expect.
Therefore, results in the left panel of Fig.~\ref{fig:noevoglx} 
corresponding to Eq.~(\ref{eq:laa})
show the maximum impact of gas depletion on the power spectrum of $B$-mode polarisation.

The contribution from galaxies appears only at small scales
and is totally overwhelmed by that from galaxy clusters.
This result tells us that we may neglect Faraday rotation from galaxies,
if most of gas is condensed onto the disk.

In Fig.~\ref{fig:cl-glxbeta},
we showed the $B$-mode power spectrum taking into account magnetic field evolution
accoding to Eq.~(\ref{eq:magneticamplitude}).
However, the magnetic field evolution,
especially in galaxy clusters, is not fully understood yet.
In some scenarios, the time scale of the evolution is shorter than the dynamical time scale
(e.g. \citet{Brandenburg-Subramanian}).
To highlight the importance of the magnetic field evolution,  
we calculate the angular spectrum
in the case of constant mangetic fields
and we plot the results in the right panel in Fig~\ref{fig:noevoglx}.
For higher $l$ modes, the main contribution comes from galaxies at redshifts higher than $z=1$.
Due to this, in the case of constant magnetic fields, 
the amplitude of the power spectrum is higher and the tail at high $l$
is also amplified.

\begin{figure}
  \begin{center}
    \includegraphics[keepaspectratio=true,height=50mm]{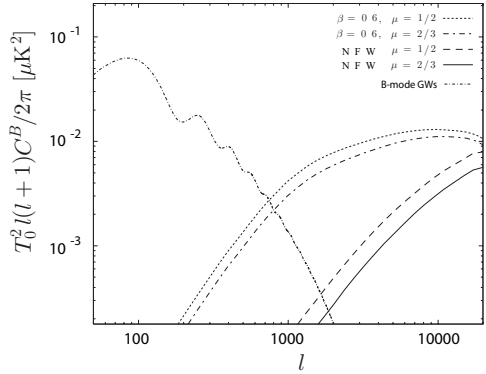}
  \end{center}
  \caption{The angular power spectra of the $B$-mode polarisation
  caused by Faraday rotation from galaxy clusters and galaxies at
  30~GHz.  The solid line is the case of the NFW profile with
  $\mu=2/3$, and the dashed line is for $\mu=1/2$.  For comparison, we
  plot the power spectra for $\beta$-profiles with $\beta =0.6$ and
  $\mu=2/3$ and $\beta =0.6$ and $\mu=1/2$ (dashed-dotted and dotted
  lines, respectively).  We also show the $B$-mode polarisation
  induced by gravitational waves with $r=0.6$.}
  \label{fig:cl-nfw}
\end{figure}

Finally, we plot the angular power spectrum of the $B$-mode
polarisation induced by Faraday rotation from galaxy clusters and
galaxies with the NFW profile in Fig.~\ref{fig:cl-nfw}.   
Here we neglect the gas depletion in galaxies and we consider the magnetic field evolution,
Eq.~(\ref{eq:magneticamplitude}).
Using the NFW
profile shifts the polarisation spectrum to smaller scales.  For $\mu=2/3$ the spectrum peaks
at $l>20000$, and the value at
$l=10000$ is about
\begin{equation}
{l(l+1) \over 2 \pi} C_l ^B \approx 4.0 \times 10^{-3} 
\left( {B_{\rm c} \over 10~\mu {\rm Gauss}} \right)^2
\left( {\nu \over 30 ~{\rm GHz} } \right)^{-4}~{\mu \rm K}^2.
\end{equation}

%\caption{The angular power spectra of the $B$-mode polarisation
% caused by Faraday rotation in galaxy clusters and galaxies.
%  For the solid line, we assume that 90 \% of gas in galaxies are condensed
%  in the disk. The dotted line shows the contribution from galaxy clusters only.
%  For comparison, we plot the power spectrum for the case without gas depletion 
%  and that due to gravitational waves with $r=0.6$ 
%  as the dashed and the dashed - dotted lines, respectively.}
%\caption{The angular power spectra of the $B$-mode polarisation
%  caused by Faraday rotation in galaxy clusters and galaxies for different 
%  evolutions of magnetic fields.  The solid line is for the model where magnetic fields
%  are constant. The dotted line corresponds to the case where magnetic fields are constant
%  and gas is depleted in galactic halos.  We
%  also plot the case where the time scale of the magnetic field evolution is 
%  the dynamical timescale as the dashed line.
%  The last line is takes into account both magnetic field evolution 
%  and gas depletion.
%  } 

%\begin{figure}
%  \begin{center}
%    \includegraphics[keepaspectratio=true,height=50mm]{glxfalling.eps}
%  \end{center}
%  \caption{The angular power spectra of the $B$mode polarisation
%  caused by Faraday rotation in galaxy clusters and galaxies.
%  For the solid line, we assume that 90 \% of gas in galaxies are condensed
%  in the disk. The dotted line shows the contribution from galaxy clusters only.
%  For comparison, we plot the power spectrum for the case without gas depletion and that 
%  due to gravitational waves with $r=0.6$ as the dashed and the dashed - dotted lines, respectively.}
%  \label{fig:glxfalling}
%\end{figure}

%\begin{figure}
%  \begin{center}
%    \includegraphics[keepaspectratio=true,height=50mm]{noevoglx.eps}
%  \end{center}
%  \caption{The angular power spectra of the $B$-mode polarisation
%  caused by Faraday rotation in galaxy clusters and galaxies for different 
%  evolutions of magnetic fields.  The solid line is for the model where magnetic fields
%  are constant. The dotted line corresponds to the case where magnetic fields are constant
%  and gas is depleted in galactic halos.  We
%  also plot the case where the time scale of the magnetic field evolution is 
%  the dynamical timescale as the dashed line.
%  The last line is takes into account both magnetic field evolution 
%  and gas depletion.
%  }
%  \label{fig:noevoglx}
%\end{figure}

%----------------------------------
%The redshift and the mass distribution of $C_l$ per $l$, shown in
%Fig.~\ref{fig:red-nfw} and Fig.~\ref{fig:mass-nfw}, indicate that the
%contribution from low redshift galaxy around $z=0.1$ is dominant

%The spectrum of
%Faraday rotation angle is decided by the low redshift galaxy around
%$z=0.1$.  Because the narrower electron density profile of the NFW
%case, the main contribution comes from the galaxies with larger mass
%than in $\beta$--profile case.  It is important to investigate the
%magnetic field strength in the galaxy with mass from $10^{12}$ to
%$10^{13} M_\odot$, in order to evaluate the spectrum of the Faraday
%rotation angle adequately.

%\begin{figure}
%  \begin{center}
%    \includegraphics[keepaspectratio=true,height=50mm]{red-nfw.eps}
%  \end{center}
%  \caption{The redshift contribution to various $l$ modes.  The short
%  dashed the solid, the long dashed and the dashed dotted lines
%  represents $l=100$, $l=500$, $l=1000$, $l=5000$ and $l=10000$,
%  respectively.  This figure is calculated in the NFW profile with
%  $\mu=2/3$.}
%  \label{fig:red-nfw}
%\end{figure}

%\begin{figure}
%  \begin{center}
%    \includegraphics[keepaspectratio=true,height=50mm]{mass-nfw.eps}
%  \end{center}
%  \caption{The mass contribution to various $l$ modes.
%  The short dashed the solid, the long dashed and the dashed dotted lines 
%  represents $l=100$, $l=500$, $l=1000$, $l=5000$ and $l=10000$, respectively.
%  This figure is calculated in the $\beta$-profile with $\beta =0.6$ and $\mu=2/3$.}
%  \label{fig:mass-nfw}
%\end{figure}

\section{discussion and summary}

%{\bf Although the amplitude for the NFW profile is lower around
%$l=10000$ than that for $\beta$-profile, the $B$-mode polarisation
%caused by Faraday rotation dominate the gravitational lensing at
%$l>6000$.}

In this study, we calculated the secondary $B$-mode polarisation
caused by Faraday rotation.  Particularly, we investigated the
dependence of the $B$-mode angular power spectrum from Faraday
rotation on the electron and the magnetic field profiles in galaxies
and galaxy clusters.  We considered both the $\beta$-profile of
electron density as well as an electron density distribution based
on the NFW dark matter profile. We modelled the magnetic field structure
in galaxies or galaxy clusters motivated by observations and further
accounted for the redshift evolution of its amplitude.

We showed that the electron and magnetic field profiles in galaxies and
galaxy clusters modify the position of the spectrum peak. The
amplitude of magnetic fields affects the amplitude of the spectrum,
mainly.  The $B$-mode polarisation from Faraday rotation at 30~GHz is
$0.01 ~{\mu \rm K}^2$ at $l=10^4$ for the $\beta$-profile,
whereas it is
$4 \times 10^{-3} ~{\mu \rm K}^2$ at the same $l$ for the NFW profile.  
We also investigated the impact of different coherence lengths on the angular power spectrum.
Small coherence length induce the depolarisation
and the peak of the power spectrum shifts to smaller scales 
and its amplitude is suppressed.

The Faraday rotation angle power spectrum as well as the $B$-mode angular
spectrum are dominated by the contribution from galaxies at redshifts
$z<1$ and with masses from $10^{11}$ to $10^{12} M_\odot$.
In other words, the detection of a secondary polarisation from Faraday
rotation would give the average strength of magnetic fields in low
redshift galaxies. 
However, the $B$-mode angular power spectrum is strongly dependent
on the evolution of magnetic fields in galaxies and galaxy clusters,
as well as on the diffuse gas fraction in galactic halos. 
If the time scale of the magnetic field evolution is shorter than the dynamical
time scale, the contribution from galaxies at redshifts higher than $z=1$ is dominant.
The power spectrum has higher amplitude and the peak is on smaller scales.  
Conversely, if the gas evolution is rapid and most of gas in galaxies
condensed onto their disks, the contribution from galaxies becomes much smaller
and can be neglected.

The contribution from Faraday rotation from clusters and galaxies
appears always small enough, on large scales, to be neglected as
compared with the primary $B$-modes from gravitational waves.  It is
also important to compare the secondary $B$-modes from Faraday
rotation with the secondary $B$-modes caused by gravitational
lensing.  The $B$-mode polarisation by Faraday rotation dominates
that due to gravitational lensing at $l>8000$ at 30~GHz.
%However, we can evaluate the $B$-mode polarisation due to
%gravitational lensing accurately  and remove that from the B-mode CMB map 
%so that the $B$-modes by Faraday rotation is detectable at even
%larger scales.  
It has been shown that gravitational lensing polarisation can be removed 
from the CMB polarisation map leaving a residual contribution of about
only 10\% \citep{hu-okamoto,hirata-seljak}.  Considering this
cleaning, the $B$-mode polarisation from Faraday rotation will exceed
the remaining signal from lensing at $l=2000$ (30~GHz), or at
$l=7000$ (100~GHz). These values are obtained in the
$\beta$-profile case with $\sigma_8=0.8$.

%In this paper, although we calculate the as the secondary
%polarisation sources, 
There are other contribution from secondary polarisation sources on small scales that we
compare to our results. One of them is polarisation from
inhomogeneous reionisation.  The coupling between the primordial CMB
temperature quadrupole anisotropy and the density fluctuations of free
electrons due to inhomogeneous reionisation produces a second
order polarisation.  Some authors estimated this polarisation by using
a semi-analytic approach \citep{liu-sugiyama,mortonson-hu}.  For
example, \citet{liu-sugiyama} gave $l(l+1)C_l^B/2 \pi \approx 10^{-5}
~{\mu \rm K}^2$ at $l \sim 10^4$.  Recently, \citet{dore-holder} have revisited this topic
using up-to-date numerical simulation.  Their results are consistent with
those of \citet{liu-sugiyama}.  
The polarisation from Faraday rotation due to the realistic magnetic fields
in galaxies and galaxy clusters
thus dominates the secondary polarisation from inhomogeneous
reionisation, at $30$~GHz.  However, Faraday rotation has a strong
dependence on frequency, $C_l\propto\nu^4$. As a result, the
polarisation by Faraday rotation becomes subdominant with respect to the
polarisation due to inhomogeneous reionisation when the frequency
exceeds $60$~GHz. Next, we considered the polarisation from
ionised gas in filaments and large scale structure (LSS).  The primary
quadrupole produces the polarisation by interaction with the free
electron gas captured in potential wells of LSS.  \citet{liu-dasilva}
have evaluated the resulting polarisation and obtained $l(l+1)C_l^B/2 \pi
\approx 10^{-3} ~{\mu \rm K}^2$ at $l \sim 10^4$.  This value is somewhat lower than the
polarisation due to Faraday rotation at $30$~GHz.  While the Faraday
rotation depends on frequency, the polarisation from LSS does not.
Therefore, although the polarisation of the Faraday rotation
dominates the polarisation of LSS at frequencies lower than 30~GHz,
the former is subdominant for $\nu<30$~GHz.

Another source of polarisation that needs to be considered is galactic foregrounds.  
On small angular scales, the
main contributions come from synchrotron emission and dust in our
Galaxy.  \citet{tucci-martinez} estimated these contributions,
focusing on the polarisation at multipoles larger than $l=1000$.  
We extrapolated their results to smaller scales and obtained,
straightforwardly, $l(l+1)C_l^B/2 \pi \approx 1 ~{\mu \rm K}^2$ 
at $l \sim 10^4$ at
$30$~GHz.  In order to detect the polarisation caused by Faraday
rotation, we need to remove the galactic foreground polarisation
carefully as it will be a major contamination. However, we have to
bare in mind that such an extrapolation relies on the assumption
that there is significant dust contribution on scales, $l\approx
10^4$. On the one hand, present observations indicate that the power spectrum of the dust
fluctuations decrease as fast as $k^{-3}$ \citep{miville2002}
suggesting a fast decrease of the dust contribution on small
scales. 
On the other hand, the contribution from synchrotron emission
on those small scales depends on the structure and amplitude of the
galactic magnetic field on the same scales. The distribution of the
magnetic fields in the Galaxy has been widely studied. 
Magnetic fields in the Galaxy will, therefore, also act as a Faraday rotation source
\citep{dineen-coles},
but their contributions need to be studied in more detail.

\section*{Acknowledgements}

We thank an anonymous referee for useful comments to improve our paper.

%\bibliography{faraday}

\begin{thebibliography}{}

\bibitem[\protect\citeauthoryear{{Athreya}, {Kapahi}, {McCarthy} \& {Bruegel}}{{Athreya}
  et~al.}{1998}]{athreya-kapahi}
{Athreya} R.~M.,  {Kapahi} B.~N.,  {McCarthy} P.~J.,    {van Breugel} W.,  
1998, A\& A, 329, 809

\bibitem[\protect\citeauthoryear{{Binney} \& {Merrifield}}{{Binney} \&
  {Merrifield}}{1998}]{binney-galactic}
{Binney}, J. J., {Merrifield}, M.,  1998, Galactic Astronomy, 1st edition, Princeton University Press


\bibitem[\protect\citeauthoryear{{Brandenburg} \& {Subramanian}}{{Brandenburg} \&
  {Subramanian}}{2005}]{Brandenburg-Subramanian}
{Brandenburg}, A., {Subramanian}, K.,  2005, Physics Reports, 417, 1


\bibitem[\protect\citeauthoryear{{Carilli} \& {Taylor}}{{Carilli} \&
  {Taylor}}{2002}]{carilli-taylor}
{Carilli} C.~L.,  {Taylor} G.~B.,  2002, Annual Review of Astronomy and
  Astrophysics, 40, 319

\bibitem[\protect\citeauthoryear{{Cavaliere} \& {Fusco-Femiano}}{{Cavaliere} \&
  {Fusco-Femiano}}{1978}]{cavaliere-fuscofemiano}
{Cavaliere} A.,  {Fusco-Femiano} R.,  1978, A\&A, 70, 677

\bibitem[\protect\citeauthoryear{{Cole} \& {Kaiser}}{{Cole} \&
  {Kaiser}}{1988}]{cole-kaiser}
{Cole} S.,  {Kaiser} N.,  1988, MNRAS, 233, 637

\bibitem[\protect\citeauthoryear{{Cooray} \& {Sheth}}{{Cooray} \&
  {Sheth}}{2002}]{cooray-sheth}
{Cooray} A.,  {Sheth} R.,  2002, Phys. Rep., 372, 1


\bibitem[\protect\citeauthoryear{{Dineen} \& {Coles}}{{Dineen} \&
  {Coles}}{2005}]{dineen-coles}
{Dineen}, P.,  {Coles}, P.,  2005, MNRAS, 362, 403


\bibitem[\protect\citeauthoryear{{Dor\'e}, {Holder}, {Alvarez}, {Iliev},
  {Mellema}, {Pen} \& {Shapiro}}{{Dor\'e} et~al.}{2007}]{dore-holder}
{Dor\'e} O.,  {Holder} G.,  {Alvarez} M.,  {Iliev} I.~T.,  {Mellema} G.,  {Pen}
  U.-L.,    {Shapiro} P.~R.,  2007, Phys. Rev. D, 76, 043002

\bibitem[\protect\citeauthoryear{{Greenfield}, {Roberts} \&
  {Burke}}{{Greenfield} et~al.}{1985}]{greenfield-roberts}
{Greenfield} P.~D.,  {Roberts} D.~H.,    {Burke} B.~F.,  1985, ApJ, 293, 370

\bibitem[\protect\citeauthoryear{{Hirata} \& {Seljak}}{{Hirata} \&
  {Seljak}}{2003}]{hirata-seljak}
{Hirata} C.~M.,  {Seljak} U.,  2003, Phys. Rev. D, 67, 043001

\bibitem[\protect\citeauthoryear{{Hu} \& {Okamoto}}{{Hu} \&
  {Okamoto}}{2002}]{hu-okamoto}
{Hu} W.,  {Okamoto} T.,  2002, ApJ, 574, 566

\bibitem[\protect\citeauthoryear{{Hummel}, {Beck} \& {Dahlem}}{{Hummel}
  et~al.}{1991}]{hummel-beck}
{Hummel} E.,  {Beck} R.,    {Dahlem} M.,  1991, A\&A, 248, 23

\bibitem[\protect\citeauthoryear{{Kamionkowski}, {Kosowsky} \&
  {Stebbins}}{{Kamionkowski} et~al.}{1997a}]{kamiokowski-kosowsky}
{Kamionkowski} M.,  {Kosowsky} A.,    {Stebbins} A.,  1997a, Phys. Rev. Lett.,
  78, 2058

\bibitem[\protect\citeauthoryear{{Kamionkowski}, {Kosowsky} \&
  {Stebbins}}{{Kamionkowski} et~al.}{1997b}]{kamiokowski-kosowski-bmode}
{Kamionkowski} M.,  {Kosowsky} A.,    {Stebbins} A.,  1997b, Phys. Rev. D, 55,
  7368

\bibitem[\protect\citeauthoryear{{Kim} \& {Kronberg}}{{Kim} et~al.}{1989}]
{kim-kromberg}
{Kim} K.-T,  {Kromberg} P.~P.,  {Giovannini} G.,    {Venturi} T.,  1989,
 Nature, 341, 720


\bibitem[\protect\citeauthoryear{{Komatsu} \& {Kitayama}}{{Komatsu} \&
  {Kitayama}}{1999}]{komatsu-kitayama}
{Komatsu} E.,  {Kitayama} T.,  1999, ApJ, 526, L1
\bibitem[\protect\citeauthoryear{{Komatsu} \& {Seljak}}{{Komatsu} \&
  {Seljak}}{2002}]{komatsu-seljak}
{Komatsu} E.,  {Seljak} U.,  2002, MNRAS, 336, 1256

\bibitem[\protect\citeauthoryear{{Kosowsky}, {Kahniashvili}, {Lavrelashvili} \&
  {Ratra}}{{Kosowsky} et~al.}{2005}]{kosowsky-kahniashvili}
{Kosowsky} A.,  {Kahniashvili} T.,  {Lavrelashvili} G.,    {Ratra} B.,  2005,
  Phys. Rev. D, 71, 043006

\bibitem[\protect\citeauthoryear{{Lewis}}{{Lewis}}{2004}]{lewis}
{Lewis} A.,  2004, Phys. Rev. D, 70, 043011

\bibitem[\protect\citeauthoryear{{Liu}, {da Silva} \& {Aghanim}}{{Liu}
  et~al.}{2005}]{liu-dasilva}
{Liu} G.-C.,  {da Silva} A.,    {Aghanim} N.,  2005, ApJ, 621, 15

\bibitem[\protect\citeauthoryear{{Liu}, {Sugiyama}, {Benson}, {Lacey} \&
  {Nusser}}{{Liu} et~al.}{2001}]{liu-sugiyama}
{Liu} G.-C.,  {Sugiyama} N.,  {Benson} A.~J.,  {Lacey} C.~G.,    {Nusser} A.,
  2001, ApJ, 561, 504

\bibitem[\protect\citeauthoryear{{Mack}, {Kahniashvili} \& {Kosowsky}}{{Mack}
  et~al.}{2002}]{mack-kahniashvili}
{Mack} A.,  {Kahniashvili} T.,    {Kosowsky} A.,  2002, Phys. Rev. D, 65,
  123004

\bibitem[\protect\citeauthoryear{{Makino} \& {Suto}}{{Makino} \&
  {Suto}}{1993}]{makino-suto}
{Makino} N.,  {Suto} Y.,  1993, ApJ, 405, 1

\bibitem[\protect\citeauthoryear{{Miville-Desch\^enes} \& {Lagache}}{{Miville-Desch\^enes} 
  et~al.}{2002}]{miville2002}
{Miville-Desch\^enes} M.-A., {Lagache} G. \& {Puget} J.-L.,
2002, A \& A, 393, 749 

\bibitem[\protect\citeauthoryear{{Mohr}, {Mathiesen} \& {Evrard}}{{Mohr}
  et~al.}{1999}]{mohr-mathiesen}
{Mohr} J.~J.,  {Mathiesen} B.,    {Evrard} A.~E.,  1999, ApJ, 517, 627

\bibitem[\protect\citeauthoryear{{Mortonson} \& {Hu}}{{Mortonson} \&
  {Hu}}{2006}]{mortonson-hu}
{Mortonson} M.~J.,  {Hu} W.,  2007, ApJ, 657, 1

\bibitem[\protect\citeauthoryear{{Murgia}, {Govoni}, {Feretti}, {Giovannini},
  {Dallacasa}, {Fanti}, {Taylor} \& {Dolag}}{{Murgia}
  et~al.}{2004}]{murgia-govoni}
{Murgia} M.,  {Govoni} F.,  {Feretti} L.,  {Giovannini} G.,  {Dallacasa} D.,
  {Fanti} R.,  {Taylor} G.~B.,    {Dolag} K.,  2004, A\&A, 424, 429

\bibitem[\protect\citeauthoryear{{Nakamura} \& {Suto}}{{Nakamura} \&
  {Suto}}{1997}]{nakamura-suto}
{Nakamura} T.~T.,  {Suto} Y.,  1997, Progress of Theoretical Physics, 97, 49

\bibitem[\protect\citeauthoryear{{Navarro}, {Frenk} \& {White}}{{Navarro}
  et~al.}{1997}]{navarro-frenk}
{Navarro} J.~F.,  {Frenk} C.~S.,    {White} S.~D.~M.,  1997, ApJ, 490, 493

\bibitem[\protect\citeauthoryear{{Ohno}, {Takada}, {Dolag}, {Bartelmann} \&
  {Sugiyama}}{{Ohno} et~al.}{2003}]{ohno-takada}
{Ohno} H.,  {Takada} M.,  {Dolag} K.,  {Bartelmann} M.,    {Sugiyama} N.,
  2003, ApJ, 584, 599

\bibitem[\protect\citeauthoryear{{Page}, {Hinshaw}, {Komatsu}, {Nolta},
  {Spergel} \& {Bennett}}{{Page} et~al.}{2006}]{page-wmap}
{Page} L.,  et al.,  2007, ApJS, 657, 1

\bibitem[\protect\citeauthoryear{{Scoccimarro}, {Sheth}, {Hui} \&
  {Jain}}{{Scoccimarro} et~al.}{2001}]{scocciarro-sheth}
{Scoccimarro} R.,  {Sheth} R.~K.,  {Hui} L.,    {Jain} B.,  2001, ApJ, 546, 20

\bibitem[\protect\citeauthoryear{{Sc{\'o}ccola}, {Harari} \&
  {Mollerach}}{{Sc{\'o}ccola} et~al.}{2004}]{scoccola-harari}
{Sc{\'o}ccola} C.,  {Harari} D.,    {Mollerach} S.,  2004, Phys. Rev. D, 70,
  063003

\bibitem[\protect\citeauthoryear{{Seljak} \& {Zaldarriaga}}{{Seljak} \&
  {Zaldarriaga}}{1997}]{Seljak-Zaldarriaga}
{Seljak} U.,  {Zaldarriaga} M.,  1997, Phys. Rev. Lett., 78, 2054

\bibitem[\protect\citeauthoryear{{Seshadri} \& {Subramanian}}{{Seshadri} \&
  {Subramanian}}{2001}]{seshadri-subramanian}
{Seshadri} T.~R.,  {Subramanian} K.,  2001, Phys. Rev. Lett., 87, 101301

\bibitem[\protect\citeauthoryear{{Sheth} \& {Tormen}}{{Sheth} \&
  {Tormen}}{1999}]{seth-tormen}
{Sheth} R.~K.,  {Tormen} G.,  1999, MNRAS, 308, 119

\bibitem[\protect\citeauthoryear{{Spergel}, {Bean}, {Dore'}, {Nolta} \&
  {Bennett}}{{Spergel} et~al.}{2006}]{spergel-wmap}
{Spergel} D.~N., et al.,
  2007, ApJS, 170, 377

\bibitem[\protect\citeauthoryear{{Subramanian}, {Seshadri} \&
  {Barrow}}{{Subramanian} et~al.}{2003}]{subramanian-seshadri}
{Subramanian} K.,  {Seshadri} T.~R.,    {Barrow} J.~D.,  2003, MNRAS, 344, L31

\bibitem[\protect\citeauthoryear{{Takada}, {Ohno} \& {Sugiyama}}{{Takada}
  et~al.}{2001}]{takada-ohno}
{Takada} M.,  {Ohno} H.,    {Sugiyama} N.,  2001, astro-ph/0112412

\bibitem[\protect\citeauthoryear{{Tashiro}, {Sugiyama} \& {Banerjee}}{{Tashiro}
  et~al.}{2006}]{tashiro-sugiyama}
{Tashiro} H.,  {Sugiyama} N.,    {Banerjee} R.,  2006, Phys. Rev. D, 73, 023002

\bibitem[\protect\citeauthoryear{{Tucci}, {Mart{\'{\i}}nez-Gonz{\'a}lez},
  {Vielva} \& {Delabrouille}}{{Tucci} et~al.}{2005}]{tucci-martinez}
{Tucci} M.,  {Mart{\'{\i}}nez-Gonz{\'a}lez} E.,  {Vielva} P.,    {Delabrouille}
  J.,  2005, MNRAS, 360, 935

\bibitem[\protect\citeauthoryear{{Varshalovich} et~al.}{{Varshalovich} et~al.}{2005}]{varshalo}
{Varshalovich} D.~A.,  {Moskalev} A.~N.,  {Khersonskii} V.K., 1988, 
{\it Quantum Theory of Angular Momentum}, (World Scientific, Singapore)


\bibitem[\protect\citeauthoryear{{Widrow}}{{Widrow}}{2002}]{widrow}
{Widrow} L.~M.,  2002, Reviews of Modern Physics, 74, 775

\bibitem[\protect\citeauthoryear{{Zaldarriaga} \& {Seljak}}{{Zaldarriaga} \&
  {Seljak}}{1997}]{zaldarriaga-seljak}
{Zaldarriaga} M.,  {Seljak} U.,  1997, Phys. Rev. D, 55, 1830



\end{thebibliography}

\appendix
\section{$B$-mode angular power spectrum}

In this appendix, we deal with the derivation of Eq.~(\ref{B--mode-result})
from Eq.~(\ref{eq:fara-eb}).
Eq.~(\ref{eq:fara-eb}) involve the integrations of the spin-weighted spherical harmonics 
in the term introduced by Faraday rotation.  
The integrations of the spin-weighted spherical harmonics are calculated
through the Clebsch-Gordan coefficients,
\begin{eqnarray}
\int d \Omega~ {}_{\pm 2}Y ^{m*} _l
Y^{m_1} _{l_1} {}_{\pm 2}Y ^{m_2} _{l_2}
& = & 
\int d \Omega~ {}_{\pm 2}Y ^{m*} _l
\sqrt{2 l_1 +1 \over 4 \pi} \sqrt{2 l_2 +1 \over 4 \pi}
\sum_{l_3 m_3} 
C^{l_3 m_3} _{l_1 m1~ l_2 m_2}
C^{l_3 \mp2} _{l_1 0~ l_2 \mp2}
\sqrt{4 \pi \over 2 l_3 +1 } {}_{\pm 2}Y ^{m_3} _{l_3}
\nonumber \\
&=&
\sqrt{{2 l_1 +1 \over 4 \pi} {2 l_2 +1 \over 2 l +1}}
C^{l m} _{l_1 m1~ l_2 m_2}
C^{l \mp2} _{l_1 0~ l_2 \mp2},
\label{eq:integral-ylm}
\end{eqnarray}
where we use the following property of the spin weighted harmonics,
\begin{equation}
{}_{s_1}Y^{m_1} _{l_1} {}_{s_2}Y ^{m_2} _{l_2}
 = \sqrt{2 l_1 +1 \over 4 \pi} \sqrt{2 l_2 +1 \over 4 \pi}
\sum_{l_3 m_3 s_3} 
C^{l_3 m_3} _{l_1 m1~ l_2 m_2}
C^{l_3 -s_3} _{l_1 -s_1~ l_2 -s_2}
\sqrt{4 \pi \over 2 l_3 +1 } {}_{s_3}Y ^{m_3} _{l_3},
\end{equation}
and the orthogonality condition
\begin{equation}
\int d \Omega~ {}_{s}Y ^{m_1*} _{l_1}
{}_{s}Y ^{m_2} _{l_2}
=
\delta _{m_1 m_2} \delta _{l_1l_2}.
\end{equation}
Substituting Eq.~(\ref{eq:integral-ylm}) to Eq.~(\ref{eq:fara-eb}), we
obtain the representation of $B$-mode polarisation from Faraday
rotation
\begin{equation}
B'_{lm}= {B}_{lm}  
-\sum_{l_1 m_1} \sum_{l_2 m_2}
[1+(-1)^{l_1+l_2-l}]
\sqrt{{2 l_1 +1 \over 4 \pi} {2 l_2 +1 \over 2 l +1}}
C^{l m} _{l_1 m_1~ l_2 m_2}
C^{l 2} _{l_1 0~ l_2 2}
\alpha_{l_1 m_1} \left({E _{l_2 m_2} -i B _{l_2 m_2}} \right),
\label{eq:b-modelm}
\end{equation}
where 
we use the following relation to obtain the last equation
\begin{equation}
C^{c \gamma } _{a \alpha ~ b \beta} = (-1)^{a+b-c} C^{c -\gamma } _{a -\alpha ~ b -\beta}.
\end{equation}

Next, we derive the $B$-mode polarisation produced by Faraday
rotation $\Delta B_{lm}$.  We assume that the rotation fields and the
primordial polarisation is statistically isotropically and uncorrelated,
\begin{equation}
\langle \alpha ^* _{l_1 m_1} E^* _{l_2 m_2} \alpha_{l_3 m_3} E _{l_4 m_4} \rangle
= \delta _{l_1 l_3} \delta _{m_1 m_3} \delta _{l_2 l_4} \delta _{m_2 m_4} 
C^\alpha (l_1) C^{E} (l_2) .
\label{eq:static-isotoropy}
\end{equation}
The angular power spectrum of the $B$-mode polarisation from Faraday rotation 
$C ^{\Delta B}(l)$ is given by 
\begin{eqnarray}
C ^{\Delta B} _l &= &\langle | \Delta B_{lm} |^2 \rangle
\nonumber \\
&=& \sum_{l_1 } \sum_{l_2 }
[(1+(-1)^{l_1+l_2-l} ) C^{l 2} _{l_1 0~ l_2 2}]^2
{2 l_1 +1 \over 4 \pi} {2 l_2 +1 \over 2 l +1}
C^\alpha _{l_1} \left (C^{E} _{l_2} -C^{B} _{l_2} \right),
\label{eq:b-polar-1} 
\end{eqnarray}
where we use the orthogonality of the Clebsch-Gordan coefficients, $
\sum_{\alpha \beta} C^{c \gamma } _{a \alpha ~ b \beta} C^{c' \gamma'
} _{a \alpha ~ b \beta} =\delta_{c c'} \delta_{\gamma \gamma'}$.
Eq.~(\ref{eq:b-polar-1}) shows that the Faraday rotation produces
$B$-mode polarisation from $E$-mode polarisation and transforms a
part of the preexisting $B$-mode polarisation into $E$-modes.

It is hard to calculate Eq.~(\ref{eq:b-polar-1}) directly.  To reduce
the amount of the calculations, we rewrite Eq.~(\ref{eq:b-polar-1})
using the following equations about the Clebsch-Gordan coefficients
\citep{varshalo},
\begin{equation}
C^{c \gamma } _{a \alpha ~ b \beta} = (-1)^{a-\alpha}\sqrt{{2c+1 \over 2b+1}}
C^{b \beta } _{c \gamma ~ a -\alpha}
= (-1)^{b+\beta}\sqrt{{2c+1 \over 2a+1}}
C^{a -\alpha } _{c -\gamma ~ b \beta}
= (-1)^{b+\beta}\sqrt{{2c+1 \over 2a+1}}
C^{a \alpha } _{ b -\beta~c \gamma },
\end{equation}
\begin{equation}
C^{c \gamma \mp1} _{a \alpha ~ b \beta}=
\sqrt{(a \mp \alpha)(a \pm \alpha+1) \over (c \pm \gamma)(c \mp \gamma+1)}
C^{c \gamma \mp1} _{a \alpha \pm 1 ~ b \beta}
+\sqrt{(b \mp \beta)(b \pm \beta+1) \over (c \pm \gamma)(c \mp \gamma+1)}
C^{c \gamma \mp1} _{a \alpha ~ b \beta \pm 1},
\end{equation}
\begin{equation}
C^{c 0} _{a 1 ~ b -1}=
\sqrt{(c +1)-a (a +1)-b (b +1) \over 2 \sqrt{a (a +1)b (b +1)} }
C^{c 0} _{a 0 ~ b 0},
\end{equation}
\begin{equation}
C^{c 2} _{a 1 ~ b 1}=
{a (a +1) [c(c +1)-a (a +1)+b (b +1)]+b (b +1)[c(c +1)+a (a +1)-b (b +1)] 
\over 2 \sqrt{a (a +1)b (b +1) c(c -1)(c +1)(c +2)}}
C^{c 0} _{a 0 ~ b 0}.
\end{equation}
After a lengthy computation, we obtain,
\begin{equation}
C^{\Delta B} _l= N_l^2 \sum_{l_1l_2}
N_{l_2}^2 K(l,l_1,l_2)^2 C^\alpha_{l_1}
{(2l_1+1)(2l_2+1)\over 4\pi(2l+1)}\left(C^{l0}_{l_10l_20}\right)^2 
\left(C^{E}_{l_2} -C^{B}_{l_2} \right),
\label{B--mode-result-2}
\end{equation}
where
\begin{equation}
N_l = (2(l-2)!/(l+2)!)^{1/2},
\label{n-factor-2}
\end{equation}
and 
\begin{equation}
K(l,l_1,l_2)\equiv -{1\over 2}\left(L^2 + L_1^2 + L_2^2 -2L_1L_2
-2L_1L +2L_1-2L_2 -2L\right),
\label{k-factor-2}
\end{equation}
with $L=l(l+1)$, $L_1=l_1(l_1+1)$, and $L_2=l_2(l_2+1)$.

To compute the Clebsch-Gordan coefficients in Eq.~(\ref{B--mode-result}),
we utilize the approximation of \citet{kosowsky-kahniashvili},
\begin{equation}
{\left(C^{c0}_{a0b0}\right)^2\over 2c+1} \approx {e\over 2\pi}
\left(1+{1\over 2g}\right)^{-2g-3/2}
\exp\left({1\over 8g}-{1\over 8(g-a)}-{1\over 8(g-b)} -
{1\over 8(g-c)}\right) \left[g(g-a)(g-b)(g-c)\right]^{-1/2},
\label{eq:clebsh-approx}
\end{equation}
This approximation is accurate to only 1\% for the worst case
$a=b=c=2$.

 \end{document}